\documentclass[sigconf]{acmart}
\AtBeginDocument{%
  }

\acmISBN{978-1-4503-XXXX-X/2018/06}

\usepackage{tabularx}
\usepackage{booktabs}
\usepackage{siunitx}
\usepackage{array}

\newcolumntype{Y}{>{\raggedright\arraybackslash}X}
\newcommand{\lcheck}{\textcolor{black!45}{\checkmark}}

\usepackage{siunitx}
\usepackage{booktabs}
\usepackage{tabularx}
\usepackage{array}
\usepackage[dvipsnames]{xcolor}
\definecolor{revisionblue}{RGB}{0, 102, 204}
\usepackage{ragged2e}   
\usepackage{enumitem}
\usepackage{graphicx}
\usepackage{subcaption}
 
\colorlet{revisionblue}{NavyBlue}

\newcolumntype{L}{>{\RaggedRight\arraybackslash}X}
\newcolumntype{P}[1]{>{\RaggedRight\arraybackslash}p{#1}}
 
 \usepackage{xcolor}

\newcommand{\systemName}{\textsc{Rehearsim}} 
\begin{document}

\title{Inspectable LLM-Simulated Participants for Qualitative Interview Rehearsal}



\author{Hangyue Zhang}
\affiliation{%
  \institution{University of Illinois Urbana-Champaign}
  \city{Champaign, Illinois}
  \country{USA}}
\email{hz85@illinois.edu}

\author{Qiran Hu}
\affiliation{%
  \institution{University of Illinois Urbana-Champaign}
  \city{Champaign, Illinois}
  \country{USA}}
\email{qiranhu2@illinois.edu}

\author{Ziyi Zhang}
\affiliation{%
  \institution{University of Illinois Urbana-Champaign}
  \city{Champaign, Illinois}
  \country{USA}}
\email{ziyiz13@illinois.edu}

\author{Hyanghee Park}
\affiliation{%
  \institution{University of Illinois Urbana-Champaign}
  \city{Champaign, Illinois}
  \country{USA}}
\email{hyanghee@illinois.edu}

\author{Yun Huang}
\affiliation{%
  \institution{University of Illinois Urbana-Champaign}
  \city{Champaign, Illinois}
  \country{USA}}
\email{yunhuang@illinois.edu}

\begin{abstract}

LLMs are increasingly used across research workflows, including research ideation, qualitative analysis, and participant simulation. Within work on participant simulation, prior studies have largely evaluated simulated participants by how realistic or human-like their responses appear, leaving less known about how researchers use and critique these materials when revising interview protocols before fieldwork. We present \systemName, an inspectable research probe that treats LLM-simulated participants as materials for methodological rehearsal rather than substitutes for human participants. \systemName\ combines literature-grounded participant construction, interactive interview rehearsal, and traceable thematic review while keeping protocol decisions under researcher control. In a study with 21 qualitative researchers, participants used simulated materials to surface assumptions and gaps, reconsider participant populations, revise target populations, and develop follow-up probes. They also identified risks, including the concealment of weak questions, dismissal of unexpected synthetic details, and interpretive anchoring from generated themes. We derive an inspectable-rehearsal framework emphasizing traceability, contestability, researcher control, and epistemic boundaries.

\end{abstract}


\begin{teaserfigure}
 \includegraphics[width=\textwidth]{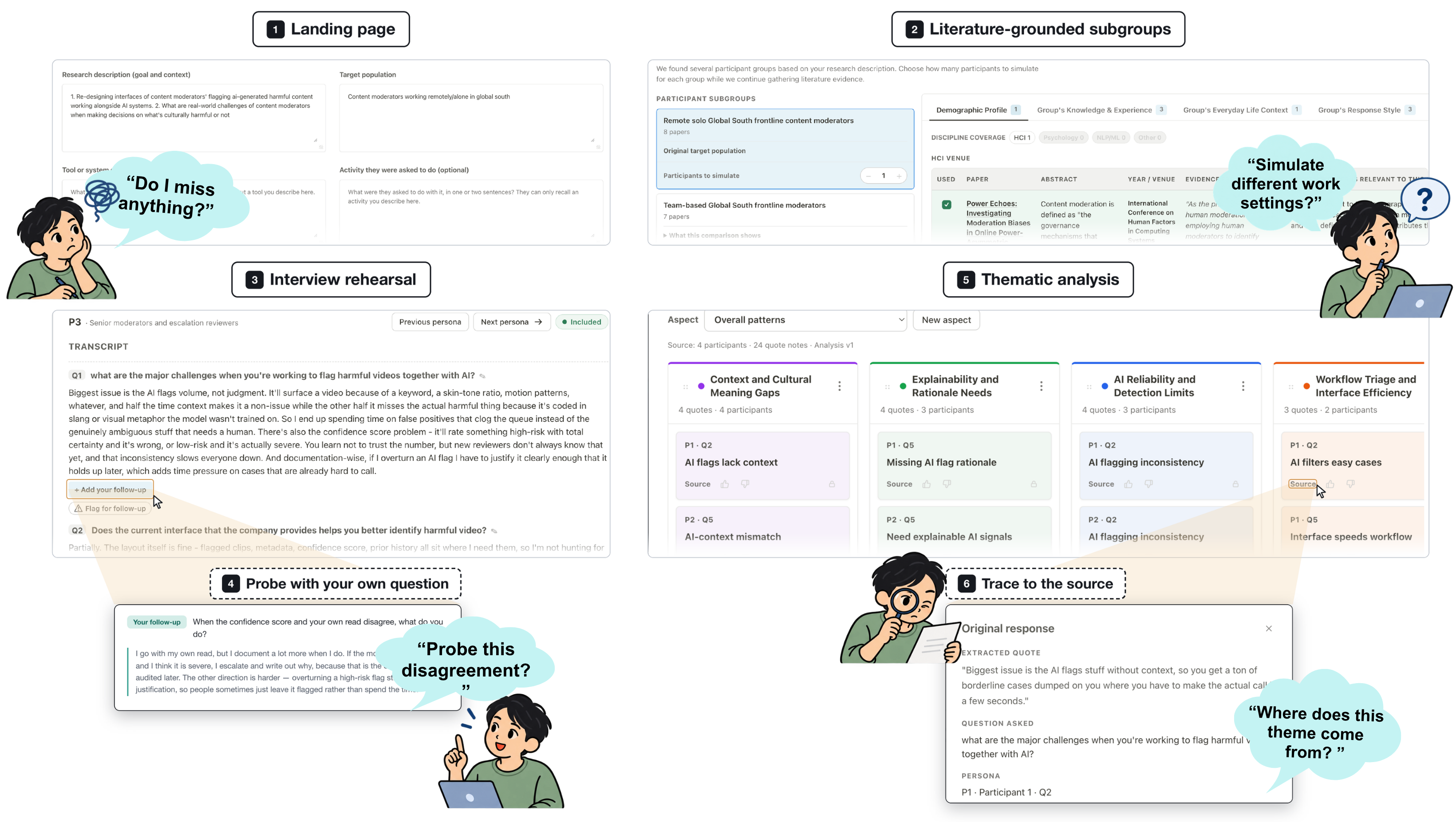}
\caption{\systemName{} experience scenario. A qualitative researcher (1) defines his study and target population on the landing page, (2) explores literature-grounded subgroups to inspect whom to include, (3) rehearses interviews with simulated participants to inspect how his questions function, (4) asks his own follow-up questions to probe specific responses more deeply, (5) reviews emerging themes in the thematic analysis dashboard, and (6) traces themes back to their source responses. All callouts are real quotes from the participant.} 
\Description{
A six-step walkthrough of the \systemName{} workflow. The researcher first enters her study context and target population, then reviews literature-grounded participant subgroups and realizes that team-based moderators may be relevant to her study. She rehearses interview questions with simulated participants and asks her own follow-up question when a response raises a new point of interest. She then reviews themes across the simulated interviews and traces individual thematic excerpts back to their original responses and interview questions. Illustrated researcher figures and thought bubbles highlight moments of reflection throughout the workflow.
}
  \label{fig:teaser}
\end{teaserfigure}


\maketitle

\section{Introduction}

Large language models (LLMs) have been increasingly used across research and scientific discovery, including literature synthesis~\cite{boyko2023,nechakhin2024}, data analysis~\cite{hollmann2023}, research ideation and hypothesis generation~\cite{kumar2024,wang2023,liu2025researchbench}, experimental planning~\cite{boiko2023,bran2023,swanson2024}, and research review~\cite{liang2023,thelwall2024,wu2026}. Among these uses, AI-simulated research participants for qualitative research have attracted particular attention, e.g., simulated domain experts to critique research ideas and refine questions~\cite{liu2025personaflow}; simulated users to generate usability feedback~\cite{xiang2024simuser,lu2025uxagent}; simulated customers and participants in shopping and human--assistant studies~\cite{sun2025shopping,liu2026humanstudy,xuan2025gidea}; and interview-informed agents for product discovery~\cite{wang2026interviewagents}.

Most prior work has focused on an \textit{output-centered} approach: it evaluates simulated personas by whether their responses appear realistic, useful, or comparable to human participants. That focus assesses simulation capability, but it says little about how researchers use and critique those outputs while designing studies. Synthetic participants can support assumption testing and iterative refinement even when imperfect~\cite{hamalainen2023evaluating,gu2025synthetic,kim2026coconstruction}, yet overly affirming interactions can also reinforce assumptions or create unwarranted confidence~\cite{gu2025synthetic}. \textit{The methodological concern is therefore not only whether simulation is accurate, but \textbf{what researchers need to inspect, question, and retain responsibility for when generated material begins to influence study design}.} 

To address this question, we iteratively designed and implemented \systemName{} as an \textbf{\textit{inspectable}} research probe, where researchers can examine and trace the intermediate representations and provenance through which generated materials are produced and organized, including participant profiles, transcripts, and theme--quote--question links.  More specifically, informed by prior work on qualitative interview design and LLM-based participant simulation, 
\systemName{} provides three components:
(1)~participant construction, through which researchers explore subgroups, select relevant literature, and inspect and edit literature-grounded profiles;
(2)~interview rehearsal, through which researchers examine simulated responses, review system-generated probes, and practice their own follow-ups; and
(3)~a thematic analysis dashboard that organizes simulated responses into an editable cross-participant view, linking themes and quotes to their source questions and interview context.
This inspectability enables researchers to question and revise generated materials while retaining control over protocol decisions; \systemName{} does not automatically rewrite the protocol. We investigate how qualitative researchers engage with \systemName{} and how simulated materials support reflection on and revision of their interview protocols. We ask two research questions:

\begin{itemize}
     \item \textbf{RQ1 (Experience)}: How do users interact with and evaluate \systemName{}?
     \item \textbf{RQ2 (Outcome)}: How does rehearsal with \systemName{} shape users’ reflections and protocol revisions?
\end{itemize}

We conducted a user study with 21 qualitative researchers, analyzing system logs, initial and revised protocols, questionnaires, think-aloud data, and follow-up interviews. 
This paper makes three contributions to HCI research on AI-supported research practice:
\begin{itemize}
  \item \textbf{\systemName{}: an inspectable research probe for methodological rehearsal.} \systemName{} makes visible how qualitative researchers examine, question, and act on LLM-simulated materials while developing interview protocols. It positions simulated profiles, responses, and themes as provisional rehearsal materials rather than evidence or substitutes for human participants, while keeping their provenance and relationships inspectable.

\item \textbf{Empirical insights into how researchers use and critically interpret simulated participants during study design.} Participants used generated materials to surface assumptions and gaps, refine research contexts and recruitment criteria, and develop follow-up probes. They also identified concrete risks: overly cooperative responses could conceal weak questions, synthetic provenance could discourage pursuit of unexpected details, and generated themes could anchor interpretation. Researchers also struggled to distinguish unfamiliar insights from model errors and cautioned against treating simulated materials as empirical evidence. Participants' perceived improvements in protocol quality only partially aligned with documented revisions, highlighting the distinction between perceived improvement and observable changes. These findings suggest that simulation can prompt methodological reflection and revision without serving as a substitute for human evidence.

  \item \textbf{An inspectable rehearsal framework for  qualitative research design.}
  Synthesizing our design and empirical findings, we characterize a suite of aspects to be inspectable when AI-generated material informs decisions about who to study, what to ask, and what a protocol elicits. The framework highlights four properties: traceability, contestability, researcher control, and an explicit epistemic boundary between rehearsal material and empirical evidence, and motivates preserving epistemic friction rather than optimizing simulations only for fluent or realistic responses.
\end{itemize}

\section{Related Work}

\subsection{LLMs in Scientific and Qualitative Research}

LLM systems for scientific research are often evaluated by output quality---predictive performance or expert agreement on well-specified tasks~\cite{dennstadt2024,lopezpineda2025,nechakhin2024}, and novelty, feasibility, or human judgment on more open-ended work~\cite{kumar2024,guo2025,liu2025researchbench,song2025}. Qualitative interview research poses a different problem: researchers must develop questions that elicit detailed accounts~\cite{castillomontoya2016preparing,adhabi2017literature,cairnslee2022enhancing} and typically refine protocols through review and piloting~\cite{castillomontoya2016preparing,willis2004cognitive}, activities that depend on methodological judgment rather than on scoring a generated artifact.

To address the workload of interpreting qualitative data, researchers have investigated LLM-assisted coding and thematic analysis~\cite{pattyn2025,kempny2026,schroeder2025,meng2025chalet}, alongside interactive tools for exploring data and developing interpretations~\cite{hayes2025,feng2026,ornelas2025,sharma2025}. To support researchers in adapting interviews to participants' responses, other systems assist with protocol navigation and follow-up generation~\cite{liu2025interviewassistant,wen2026interflow,zhang2025followup,zhang2026}. Limited access to participants can also constrain early feedback and protocol testing, motivating interest in LLM-based simulated participants for research preparation~\cite{kapania2025simulacrum}. Across research and design, simulations have been explored for ideation and design feedback~\cite{liu2025personaflow,choi2025proxona,gu2025synthetic,deep2025interactivepersonas}, diversifying synthetic populations, modeling user behavior, and evaluating usability~\cite{ren2024bases,xiang2024simuser,lu2025uxagent,sun2025shopping}, and generating responses to interviews, surveys, and product concepts~\cite{arora2025hybrids,saini2026contextualization,wang2026interviewagents}. These applications offer material for preliminary exploration without repeated participant recruitment. However, existing work primarily focuses on simulation fidelity, using information about target populations and individuals to ground simulations and to evaluate their correspondence with human responses~\cite{truss2026personacite,park2026selfreports,tan2025personabench,zhu2026realusersim}. Consequently, less is known about how materials generated through LLM-based simulation support methodological reflection before human data collection begins. 

\subsection{Inspectability and Control of Generated Materials}

Inspectability concerns whether people can examine how an interpretation was produced and where judgment entered~\cite{vermeeren2008usability}. For AI systems, exposing why an output occurred is not the same as letting users shape what happens next~\cite{westphal2023decision}. Controls that keep generated suggestions open to rejection, revision, or selective uptake can support engagement without requiring users to accept the output as a whole~\cite{dietvorst2018overcoming,cheng2023overcoming,fink2024let,schaap2024abc}.

LLM systems have begun to make alternatives available for that kind of scrutiny. Contrasting cases help users compare candidate solutions~\cite{naik2025providing}; generated counterarguments prompt reconsideration of recommendations~\cite{chiang2024enhancing}; and deliberative dialogue structures disagreement and revision~\cite{ma2025towards}. These designs treat model contributions as subject to human evaluation~\cite{woodruff2026epistemic,vendrell2026scaffolding}. They typically attach inspectability to a recommendation or a decision, however, rather than to generated \emph{research materials}---profiles, responses, and themes---that may influence who is studied and what is asked before any human data exist.

\section{\systemName{}}

\begin{figure*}[htbp]
  \centering
  \includegraphics[width=\linewidth]{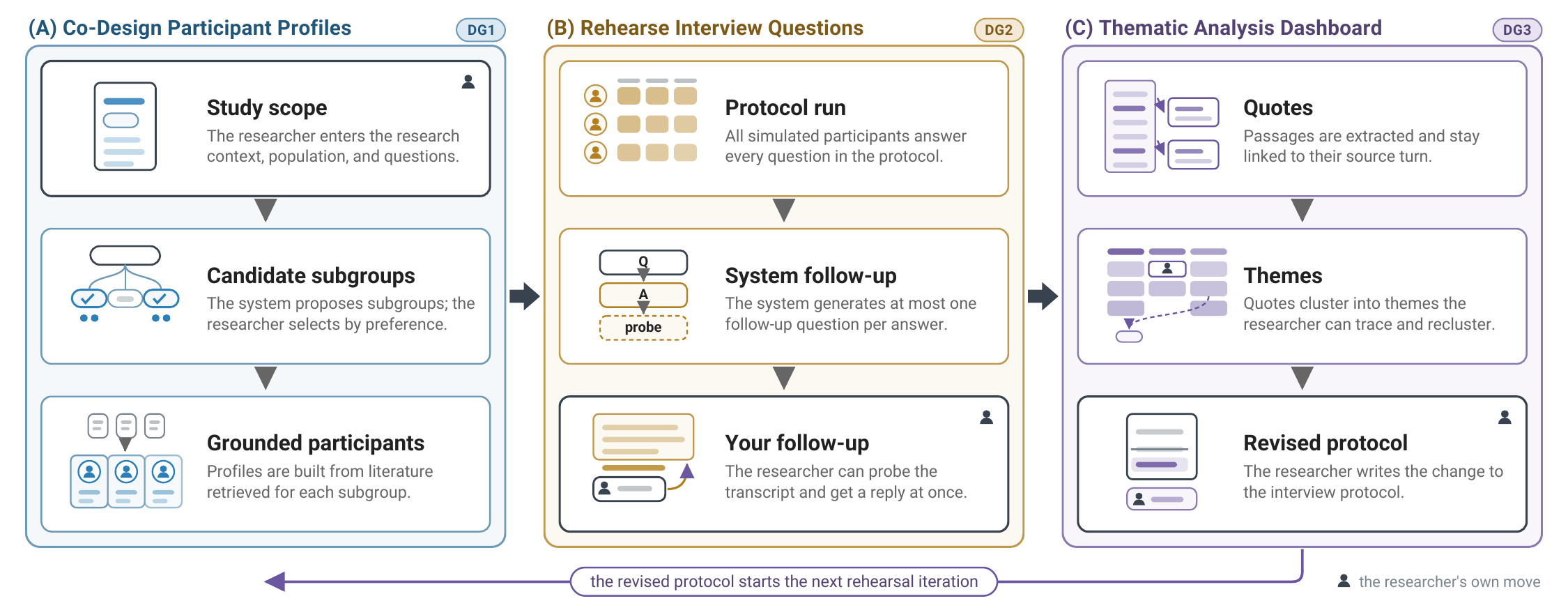}
  \caption{\systemName\ workflow. (A)~The researcher and \systemName\ co-design literature-grounded profiles: a target population is expanded into subgroups, the researcher chooses which enter, and each profile is built from literature retrieved for that subgroup. (B)~Those participants answer the protocol; the system may add one follow-up per question, and the researcher can probe further in the transcript. (C)~A dashboard clusters responses into themes that remain traceable to their source turns. The revision at the end of (C) is written by the researcher.}
  \Description{A three-panel diagram with arrows connecting the panels in order. Panel A, Co-Design Participant Profiles, holds three cards linked by downward arrows: a study-scope form, a population chip branching into three subgroup chips of which two are checked, and three participant cards fed by document icons. Panel B, Rehearse Interview Questions, holds a grid of participants by questions, a question-answer-probe chain whose probe is dashed, and a transcript with a researcher input box whose reply is appended to the transcript. Panel C, Thematic Analysis Dashboard, holds quotes lifted out of a transcript into cards, quote cards grouped into three theme columns with a dashed line tracing one back to its source, and a protocol sheet with one line struck through and one line added above a note field. An arrow beneath the three panels returns to the start of the next rehearsal iteration.}
  \label{fig:overview}
\end{figure*}

\subsection{Design Goals (DGs)}\label{sec:design-goals}

\systemName\ draws on prior research on qualitative interview design and LLM-based participant simulation. We also conducted pilot studies with six qualitative researchers to refine the system design and assess the feasibility of the study workflow.



\textbf{DG1: Surface Population Assumptions through Participant Construction.}
Prior work on LLM-based participant simulation has identified limitations in how models represent human experiences. Generated personas may reproduce demographic stereotypes~\cite{cheng2023marked}, while simulated responses may misrepresent identity groups and fail to capture within-group variation~\cite{wang2025misportray}. These concerns motivated us to make the construction of simulated participants open to researcher scrutiny. \systemName{} grounds participant construction in relevant literature, allows researchers to examine and revise proposed subgroups and profiles, and preserves traceable links between supporting literature and the profile attributes derived from it.

\textbf{DG2: Support Reflection and Revision through Interview Rehearsal.}
Effective interview design requires alignment between research objectives and interview questions, as well as attention to how different participants may interpret and respond to those questions~\cite{castillomontoya2016preparing,yeong2018interview}. Feedback and piloting can reveal problems in question wording, coverage, sequencing, and probing, yet access to suitable participants can limit such testing before fieldwork~\cite{castillomontoya2016preparing,roberts2020,kapania2025simulacrum}. We therefore designed \systemName{} to let researchers rehearse their interview protocols through simulated interactions, compare how questions elicit responses across participants, and practice follow-up probing while retaining control over whether and how the protocol is revised. 

Guided by DG1 and DG2, we developed an initial prototype in which researchers could select literature used to construct simulated participant profiles, rehearse their interview questions with those participants, and review codes and themes generated from the simulated responses. We conducted an initial pilot study with four experienced qualitative researchers (R1--R4) to gather feedback on this workflow. Participants valued the literature-grounded profiles and simulated responses, but identified several needs for refinement: (1) comparing meaningfully different participant groups to reconsider the scope of the target population (R2, R4); (2) increasing variation across simulated responses to reduce response homogeneity (R3, R4); (3) allowing researchers to adapt questions and ask their own follow-ups in response to simulated participants (R4); and (4) strengthening connections between generated themes and the underlying interview material so that researchers could assess how those interpretations were formed (R1, R2). In response, we added \textbf{DG3: Supporting Evaluation of Thematic Analysis  through Traceable Interpretations.} Enable researchers to trace generated themes back to their source responses and interview context, compare them with their own reading, and assess whether the interpretations are supported by the underlying simulated material.

\subsection{User Interaction and Workflow}
\label{sec:modules}

\systemName{} supports a three-stage rehearsal workflow corresponding to DG1--DG3. As shown in Fig.~\ref{fig:overview}, researchers first construct literature-grounded simulated participants, then rehearse their interview protocol through simulated interactions, and finally review a traceable thematic synthesis of what the protocol appears to elicit. 

\begin{figure*}[htbp]
  \centering
  \includegraphics[width=\linewidth]{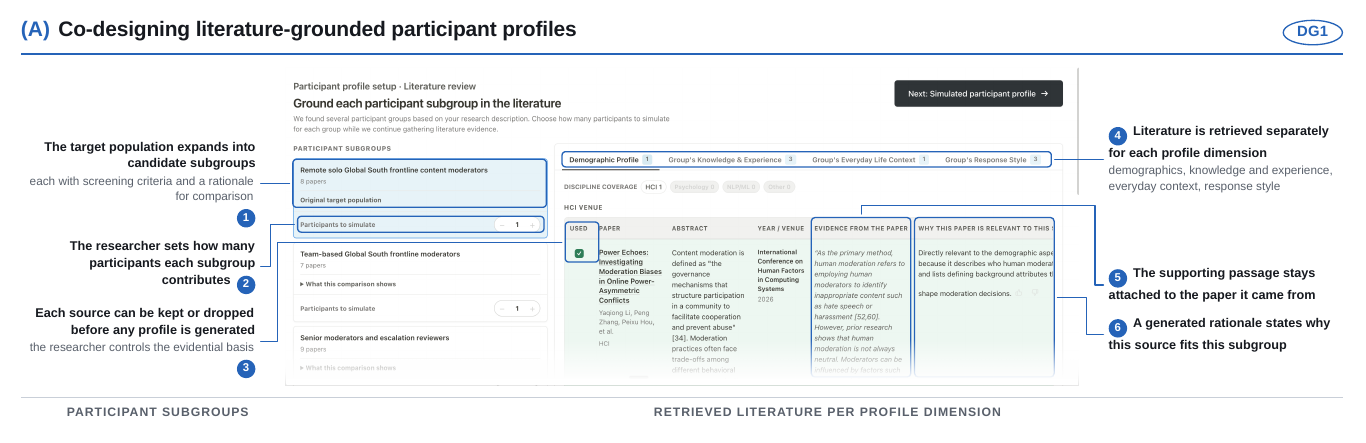}
  \caption{Literature-grounded participant construction in \systemName. (1)~The target population is expanded into candidate subgroups with screening criteria and comparison rationales. (2)~Researchers set each subgroup's representation. (3)~Retrieved sources can be inspected and removed before profile generation. (4)~Literature is retrieved separately across four profile dimensions. (5)~Passages remain traceable to source papers, and (6)~generated rationales explain each source's relevance to the subgroup.}
  \Description{A screenshot of \systemName’s participant-profile construction interface with six numbered annotations. The left side lists candidate subgroups derived from the researcher’s target population and includes controls for selecting the number of simulated participants and retaining or removing literature sources. The right side displays retrieved literature organized by four profile dimensions: demographics, knowledge and experience, everyday context, and response style. Each literature entry includes paper metadata, a supporting passage, and a generated explanation of why the paper is relevant to the subgroup. The annotations identify subgroup expansion, participant allocation, source selection, dimension-specific retrieval, passage-to-paper traceability, and relevance rationales.}
  \label{fig:dg1}
\end{figure*}

\subsubsection{Literature-Grounded Participant Profiles (\textbf{DG1})}
\label{sec:participant-construction}

As shown in Fig.~\ref{fig:dg1}, \systemName\ expands the stated target population into candidate subgroups, each with a description, screening criteria, and a rationale for comparison. Researchers select which subgroups enter rehearsal and how many simulated participants each contributes. 
Profiles are modeled along four dimensions used in prior simulated-participant work: demographic background~\cite{doudkin2025synthetic,sun2025shopping,liu2026humanstudy}, knowledge and experience~\cite{teutloff2025synthetic,truss2026personacite,shin2024understanding}, everyday context~\cite{moon2024virtual,makridis2025virtualxai,liu2026humanstudy}, and response style~\cite{ahmad2025simulating}. For each selected subgroup, \systemName\ retrieves literature on these dimensions and shows, for each paper, the abstract, venue, a relevant passage, and a rationale for inclusion. Researchers can keep or drop sources before generation, then inspect and edit the resulting profiles.

\subsubsection{Interview Rehearsal and Follow-Up Practice (\textbf{DG2})}

\begin{figure*}[htbp!]
  \centering
  \includegraphics[width=\linewidth]{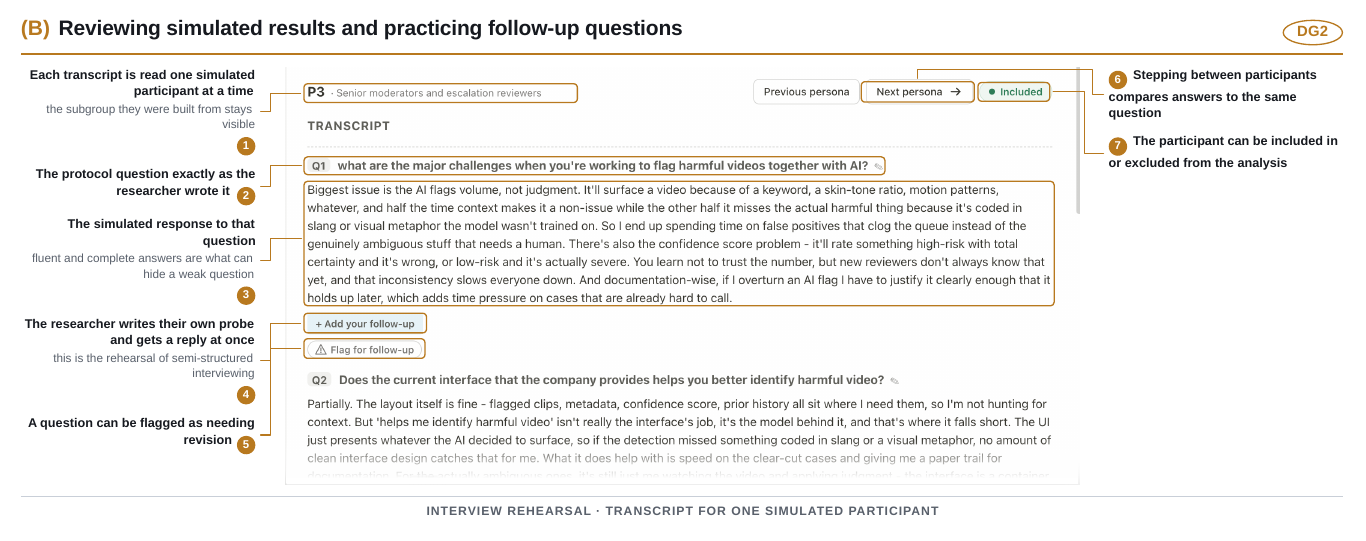}
  \caption{Interview rehearsal and follow-up practice in \systemName. (1)~Researchers review one simulated transcript at a time with the participant's subgroup visible. (2)~Each protocol question is shown as written, followed by (3)~the simulated response. (4)~Researchers can inspect suggested probes, write their own follow-ups, and receive immediate replies. (5)~Questions can be flagged for later revision. (6)~Researchers can move between participants to compare how the same questions unfold, and (7)~include or exclude participants from later analysis.}
  \Description{A screenshot of \systemName’s interview-rehearsal interface with seven numbered annotations. The central panel displays the transcript for one simulated participant, with the participant identifier and subgroup shown at the top. Each protocol question appears in sequence together with the simulated participant’s response. Controls below responses allow the researcher to add a follow-up question or identify an opportunity for further probing, and protocol questions can be flagged for revision. Navigation controls at the top allow researchers to move between simulated participants and compare their responses, while an inclusion control determines whether a participant is retained for subsequent analysis. The annotations highlight participant context, original protocol questions, simulated responses, follow-up practice, question revision, cross-participant comparison, and participant inclusion.}
  \label{fig:dg2}
\end{figure*}

\label{sec:rehearsal}
As shown in Fig.~\ref{fig:dg2}, once participant profiles are selected, 
\systemName\ runs the interview protocol with the simulated participants. Researchers can review one transcript at a time or compare how the same question unfolds across participants. 
Semi-structured interviewing depends on probing, clarification, and adaptation~\cite{dejonckheere2019semistructured,knott2022interviews,liu2025envisioning,wen2026interflow}. To rehearse that work, researchers can write their own follow-ups and receive immediate replies and flag questions for later revision.

\subsubsection{Traceable Thematic Review (\textbf{DG3})}
\label{sec:sensemaking}

\begin{figure*}[htbp]
  \centering
  \includegraphics[width=\linewidth]{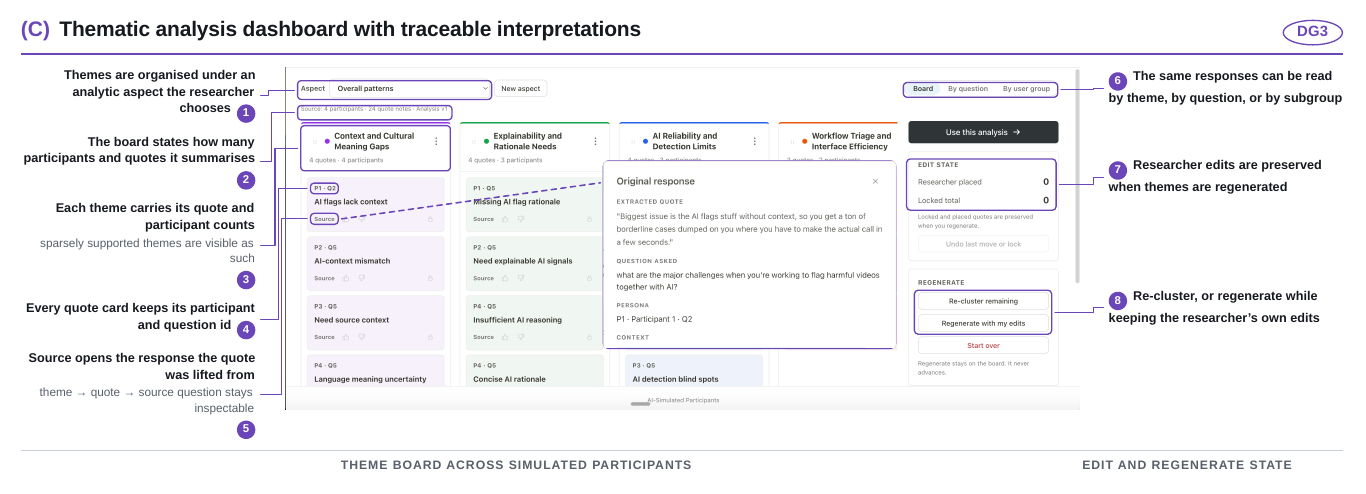}
  \caption{Thematic analysis dashboard with traceable interpretations in \systemName. (1)~Researchers organize the analysis under a chosen analytic aspect. (2)~The dashboard shows how many participants and quote notes are in the current summary. (3)~Each theme shows supporting quote and participant counts. (4)~Quote cards retain participant and source-question identifiers, and (5)~the source view links each quote to the original response. (6)~Material can be inspected by theme, question, or subgroup. (7)~Researcher edits persist across regeneration, and (8)~themes can be re-clustered or regenerated while preserving those edits.}
  \Description{A screenshot of \systemName’s thematic analysis dashboard with eight numbered annotations. The main view presents a board of themes organized under a researcher-selected analytic aspect. Each theme column contains quote cards summarizing simulated responses, along with counts showing how many quotes and participants support the theme. Each quote card retains its participant identifier and source question, and selecting a source opens a detailed view of the original response from which the quote was extracted. The interface also provides alternative ways to inspect the same material, including views by theme, by question, and by subgroup. On the right, controls allow researchers to preserve their edits when regenerating themes and to re-cluster or regenerate the thematic organization while keeping those edits. The annotations highlight researcher-selected analytic framing, summary counts, theme support, quote traceability, source inspection, alternative organizational views, edit preservation, and regeneration controls.}
  \label{fig:dg3}
\end{figure*}

To support cross-interview inspection without requiring researchers to manually synthesize every simulated transcript, \systemName{} generates a provisional thematic overview of the simulated responses. This synthesis serves as a rehearsal aid rather than as a substitute for researcher-led qualitative analysis. Its purpose is to help researchers examine what their current protocol appears to elicit across simulated participants. As shown in  Fig.~\ref{fig:dg3}, \systemName\ organizes quotes into themes using inductive and deductive coding. Each quote stays linked to its participant and source question, enabling themes and codes to be traced back to the interview turn that produced them.
Researchers can rename or merge themes, move quotes, and add their own analytic cuts. 

\subsection{System Implementation}
\label{sec:implementation}

The frontend is implemented with React~19 and TypeScript; the backend uses Python FastAPI. Study configurations, profiles, transcripts, and analysis results are stored in Neo4j. Pipeline stages use GPT-5.5, Asta~\cite{asta}, Claude Sonnet~5, and Gemini~3.5 Flash. Appendix A summarizes the prompt contracts; complete prompts are provided in the supplementary material.

\subsubsection{Subgroup Generation}

Given the research context and stated population, GPT-5.5 proposes five to eight candidate subgroups using purposive maximum-variation sampling~\cite{patton2015qualitative}: dimensions such as prior experience, institutional context, access, or resources, including both variation inside the stated population and adjacent roles that may shape the phenomenon. Because candidates can overlap, each subgroup has inclusion and exclusion criteria; a person who could fit more than one is assigned to a single subgroup or the overlap is treated as a within-subgroup screen. In each study iteration, every simulated participant belongs to exactly one subgroup.

\subsubsection{Literature Retrieval and Profile Generation}

For each selected subgroup, \systemName\ retrieves literature on the four profile dimensions in Section~\ref{sec:participant-construction}. Queries are built from the research context, subgroup definition, and profile dimension. Interview questions are excluded from retrieval, so the retrieved evidence is not conditioned on the protocol being rehearsed. For each query, Asta searches the Ai2 Scientific Corpus, a large-scale scientific literature graph containing metadata for over 225 million papers and a full-text index of over 12 million publications \footnote{Asta Scientific Corpus Tool: \url{https://allenai.org/asta/resources/mcp}}. Asta returns relevant paper metadata and ranked text snippets from the indexed literature. GPT-5.5 converts the retrieved snippets into evidence records (finding, supporting passage, sample, relevance) using extraction prompts adapted from Wright et al.~\cite{wright2025uea}. Passages are checked against the retrieved snippets so they remain traceable to their source papers.

Profiles are then generated from the subgroup definition and that evidence, adapting SPASM~\cite{luo2026spasm} and SPeCtrum~\cite{lee2025spectrum}: structured attributes plus a first-person background. Participants in the same subgroup receive different evidence subsets to introduce within-group variation. To preserve consistency, each profile is fixed for the rest of the iteration.

\subsubsection{Simulated Interview Generation}

To reduce homogeneity and sycophancy in single-model panels~\cite{anderson2024homogenization,sharma2024sycophancy}, participants are split between GPT-5.5 and Claude Sonnet~5 and keep that model for profile and interview turns.

Each interview has three roles. The participant agent answers in first person from its profile, with a prompt adapted from persona-driven role-play~\cite{cai2026thinkpersona,luo2026spasm,lee2025spectrum,hong2025sycon}. The interviewer agent presents protocol questions and may add at most one follow-up when a response is ambiguous, uncertain, underexplained, or inconsistent~\cite{dejonckheere2019semistructured,rubin2012qualitative}. Researchers can also insert their own follow-ups (Section~\ref{sec:rehearsal}). A checker adapted from CheckEval~\cite{lee2025checkeval} flags answers that miss the question, invent unsupported personal detail, or contradict earlier turns. Interviews run in parallel; turns within an interview are sequential. Each turn receives the fixed profile, response style, current question, and interview history.

\subsubsection{Thematic Analysis}

GPT-5.5 extracts quotes relevant to the research question using a query-focused prompt adapted from Wright et al.~\cite{wright2025uea}. Each quote keeps its participant, subgroup, question, and transcript location. Coding follows HICode~\cite{zhong2025hicode} and SCALE~\cite{zhao2025scale}: an inductive codebook is built from the quotes, each quote is coded against the research question, and codes are grouped into themes. Researcher follow-ups are included in the next analysis pass. Theme--quote links are what Section~\ref{sec:sensemaking} exposes in the dashboard.

\section{User Study }

\begin{figure*}[htbp]
  \centering
  \includegraphics[width=1\linewidth]{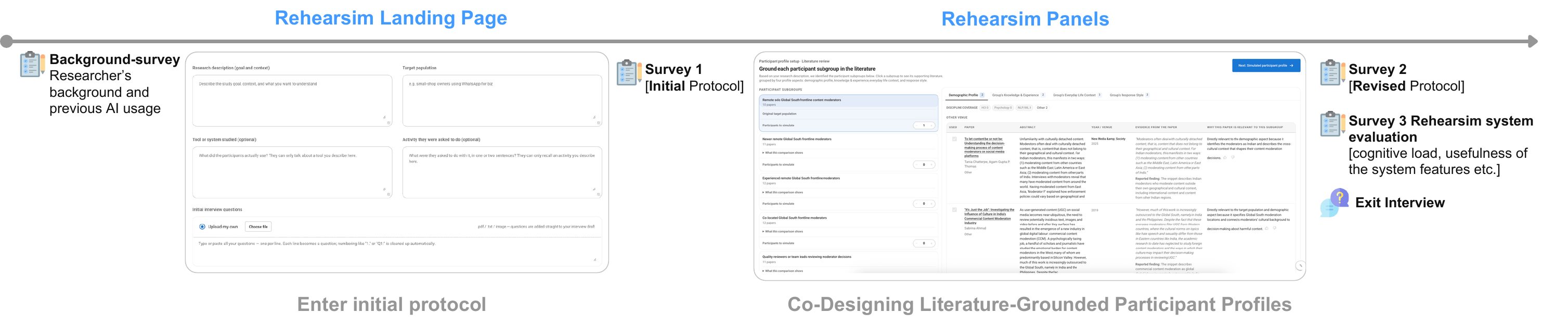}
  \caption{Participants provided an initial protocol to \systemName{}, explored the system while thinking aloud, revised their protocol, and completed surveys at different stages of the study.}
  \Description{A left-to-right diagram of the study procedure. Participants first complete a background survey and enter their research context, target population, and initial interview protocol on the \systemName\ landing page. They then complete Survey 1 to evaluate the initial protocol and use the \systemName\ panels to construct literature-grounded participant profiles, rehearse interviews, and review generated analyses. After revising their protocol, participants complete Survey 2 to evaluate the revised protocol, Survey 3 to evaluate the system, and an exit interview.}
  \label{fig:study-procedure}
\end{figure*}

We conducted a user study to evaluate \systemName{} as a tool for qualitative interview protocol rehearsal. The study focused on understanding how researchers engaged with and evaluated \systemName{} during rehearsal, and how the rehearsal process shaped their reflections and protocol revisions.


\subsection{Participants}

Our study was approved by the university's IRB. We conducted pilot studies with six qualitative researchers. Four (R1–R4) provided feedback to refine the system design, study procedure, and materials. The other two assessed the feasibility of the study workflow and are also included in the main-study sample as P1 and P2. We recruited 21 (12 female, 9 male) participants with qualitative research experience through snowball sampling and university-affiliated bulletin boards. Participants included Master's and PhD students, postdoctoral researchers, faculty members, and UX researchers from HCI, social science, education, health, communication, design, and related fields. Six participants (28.6\%) had conducted 1--2 qualitative studies, nine (42.9\%) had conducted 3--5, and six (28.6\%) had conducted six or more. Participants who reported publication counts had published an average of 5.5 qualitative research papers, with publications appearing in venues such as CHI, CSCW, UIST, IUI, iConference, AEJMC, and NCA. All participants had used generative AI in their research, most commonly for writing or literature reviews (17, 81.0\%), developing interview questions (16, 76.2\%), brainstorming research questions and refining study protocols (15 each, 71.4\%), and qualitative data analysis (13, 61.9\%); three (14.3\%) had used it to simulate participants, users, or personas. The background information of participants is shown in Appendix \ref{sec:participants}. Each session lasted approximately 1.5--2 hours, and participants received \$20 USD per hour.

\subsection{Procedure} 
Fig. ~\ref{fig:study-procedure} illustrates the study procedure. The study began with collecting participants' background information, including their qualitative research experience and prior use of AI in research. The researchers then introduced the study purpose, explained the tasks, and walked participants through the system tutorial. Participants were encouraged to bring an ongoing study at an early stage or brainstorm a study idea during the session. They entered their study context and goals, target population, and initial interview protocol into \systemName. They then completed a pre-survey capturing their perceptions of the proposed protocol (survey 1, initial protocol).

Participants were then guided through \systemName{}'s features and asked to consider potential revisions to their interview protocols based on the generated material. Throughout system use, participants were encouraged to think aloud, allowing the researchers to observe and record their interactions and thoughts. After exploring the system, participants were asked to revise their initial protocols and explain their decisions. They then completed exit surveys assessing their perceptions of the revised protocol (survey 2, revised protocol) and their experience with the system (survey 3, \systemName\ system evaluation). Finally, participants were interviewed to further explore their overall experiences and gather feedback on their protocols and the system features.

\begin{figure*}[t]
    \centering
    \includegraphics[width=\linewidth]{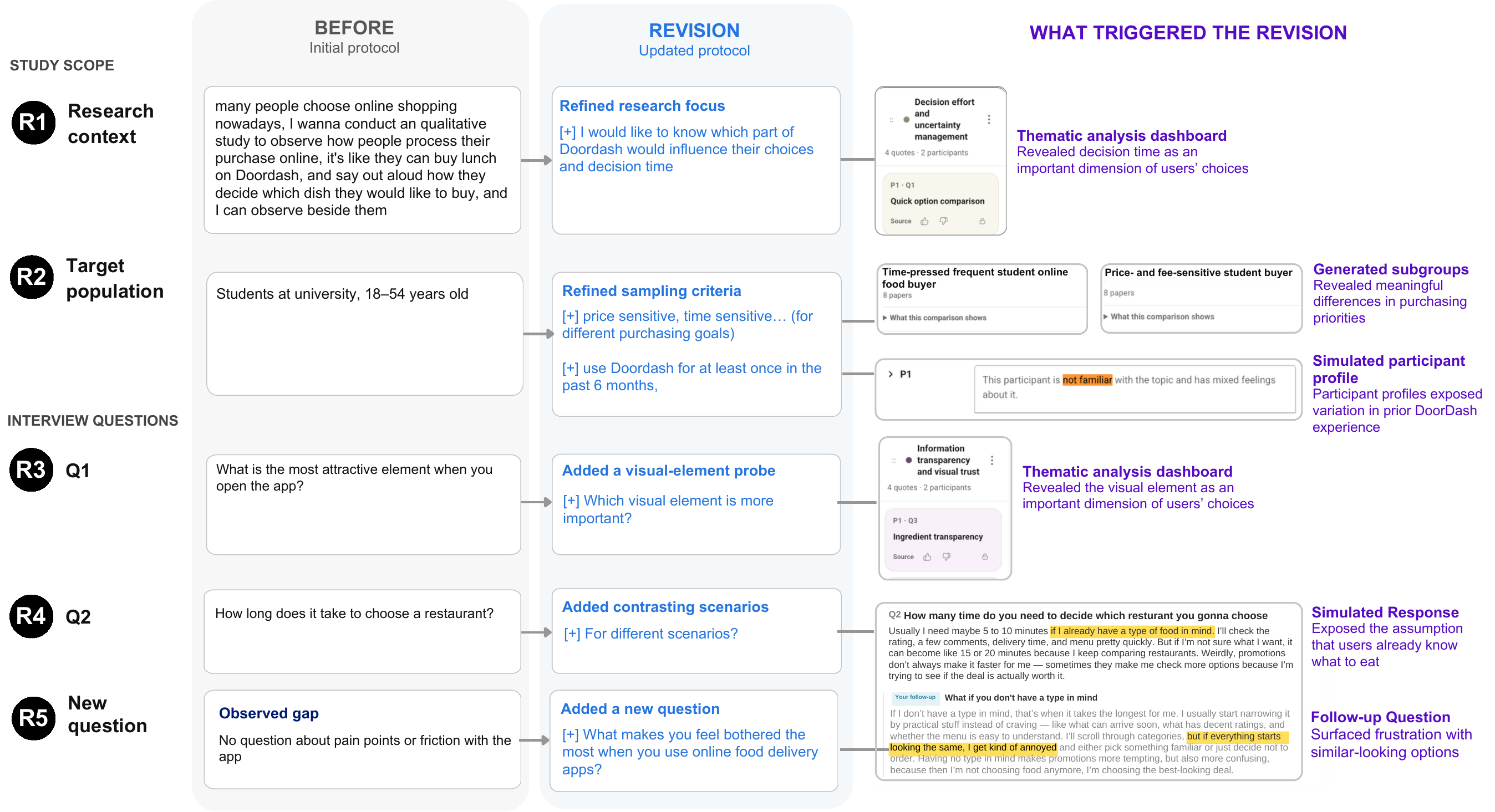}
    \caption{P4’s protocol revisions through \systemName. The figure traces five revisions from P4’s initial protocol (left) to the updated protocol (middle), together with the system outputs and interactions that triggered each change (right). Generated subgroups and participant profiles revised the target population and inclusion criteria; thematic analysis highlighted visual cues and decision time as relevant dimensions; a simulated response exposed an assumption that users already knew what they wanted to eat; and P4’s own follow-up surfaced frustration with similar-looking options, motivating a new interview question.}
\Description{
A three-column diagram illustrating how P4 revised a qualitative interview protocol while using \systemName. The left column shows the initial protocol, the middle column shows the revised protocol, and the right column shows the system outputs or interactions that triggered each revision. Five revisions are shown. First, the research context was expanded to examine which aspects of DoorDash influence users' choices and decision time, prompted by themes related to decision effort and uncertainty. Second, the target population was revised to include price- and time-sensitive students and recent DoorDash users, prompted by generated subgroups and simulated participant profiles. Third, a probe about which visual element is most important was added after the thematic analysis highlighted visual elements in participants' choices. Fourth, the question about restaurant decision time was revised to compare different scenarios after a simulated response exposed the assumption that users already knew what they wanted to eat. Fifth, a new question about frustrations with food-delivery apps was added after P4's follow-up question surfaced annoyance with similar-looking options.
}
    \label{fig:p4-revisions}
\end{figure*}

\subsection{Data Collection and Analysis}

We collected system logs, initial and revised research protocols, questionnaire responses, think-aloud data, and follow-up interview transcripts to examine participants' protocol revisions, perceived outcomes, and experiences with \systemName.

\subsubsection{Questionnaires.}
\label{sec:questionnaire}
The questionnaires covered three areas. First, a background questionnaire captured participants' qualitative research experience and prior use of AI for research. Second, Surveys 1 and 2 assessed participants' perceptions of their protocols before and after using \systemName. Protocol quality items covered clarity of the research scope and target population, coverage, clarity, depth, breadth, methodological rigor, and feasibility. These items were adapted from methodological guidance on developing and refining semi-structured interview protocols~\cite{castillomontoya2016preparing, kallio2016systematic, yeong2018interview, dejonckheere2019semistructured}. Additional items assessed participants' perceived ability to identify assumptions and contextual gaps, identify concrete revisions, prioritize revisions, and revise their protocols with confidence. Third, Survey 3 assessed perceptions of \systemName{}'s features across five dimensions: enjoyment of use~\cite{davidson2023development}, perceived usefulness~\cite{davis1989perceived}, trust in the feature's suggestions and decisions~\cite{everard2005presentation}, support for critical reflection, and perceived contribution to protocol revision. The details of surveys are shown in Appendix \ref{survey}.

For all these measures, we conducted a Shapiro-Wilk test and found that the data were not normally distributed. We used Wilcoxon signed-rank tests to compare paired pre- and post-use ratings and reported rank-biserial correlations as effect sizes. We used Kendall's $\tau_b$ to explore associations among participants' backgrounds, perceptions of system features, and extent of protocol revision.

\subsubsection{System Logs and Protocol Revisions.}
We recorded participants' initial inputs, including their research goals and context, target population, the tool or system under study, planned participant activities, and interview questions. We also logged subgroup selections, the number of simulated participants, edits to simulated participant profiles, follow-up questions, and theme clicks. At the end of each session, we collected the revised protocol.

We summarized feature usage and compared each participant's initial and revised protocols. We examined these changes alongside system logs and think-aloud data to understand how participants used generated materials to inform their revisions. To characterize what aspects of protocol design participants revised, two researchers independently coded revision moves using the eight protocol-quality dimensions assessed in our surveys. Coding drew on both the protocol artifacts and participants' think-aloud explanations. Each revision could be associated with multiple dimensions. Disagreements were resolved through discussion and consensus. 

To examine the alignment between perceived changes in protocol quality and documented revisions, we compared participants' pre- and post surveys with coded protocol revisions. For each participant, we identified which dimensions showed changes in their self-assessment ratings and which were represented in their coded revisions. We then assessed the agreement between these two sets of dimensions using percentage agreement and Cohen's $\kappa$ across 168 comparisons (21 participants $\times$ 8 dimensions). At the participant level, we used Spearman's $\rho$ to examine whether participants who reported changes across more dimensions also made revisions spanning more dimensions.

\subsubsection{Think-Aloud and Interviews.}
We recorded participants' think-aloud comments as they explored \systemName\ and revised their protocols, prompting them to explain their interpretations of generated materials and reasons for accepting, modifying, or rejecting suggestions. Follow-up semi-structured interviews covered three areas: (1) the usefulness of individual features and their role in protocol revision; (2) the overall experience, perceived changes to the protocol, and desired improvements; and (3) comparisons with participants' usual research practices, including their use of existing AI platforms. We also asked how simulated participants shaped their thinking about target populations, interview questions, and underlying assumptions, and what benefits and risks they perceived in using simulation for research preparation.

We conducted a thematic analysis of think-aloud and interview transcripts to examine how participants used and evaluated \systemName. First, two researchers independently conducted an initial round of open coding on a subset of transcripts to identify salient patterns in participants' reflections, revisions, perceived benefits and risks, and desired improvements. Codes captured how participants interpreted simulated materials, how specific system features shaped their judgments, and how they described changes to their protocols. The two researchers then compared codes and iteratively refined the codebook through discussion, consolidating overlapping codes and clarifying definitions. The finalized codebook was applied to the full dataset, with disagreements resolved through discussion and consensus. We examined coded accounts alongside recorded system interactions and participants' protocol revisions to contextualize how particular features were used and how reported reflections related to observed changes. Finally, we grouped coded excerpts into broader thematic categories aligned with our research questions.

\section{Findings}


In this section, we first illustrate how participants used \systemName\ through a user journey in Section~\ref{sec:user_journey}. We then present findings on how participants used the system and on perceived benefits and risks (RQ1; Section~\ref{sec:rq1}), and on protocol revisions and self-reported changes in protocol quality and revision readiness (RQ2; Section~\ref{sec:rq2}).

\subsection{Sample User Journey}
\label{sec:user_journey}
As shown in Fig. ~\ref{fig:p4-revisions}, P4 was designing a qualitative study on how university students make food-ordering decisions on DoorDash, and her protocol evolved across \systemName's stages. Initially, P4 defined her target population as university students aged 18--54. Subgroup recommendations prompted her to add price and time sensitivity into recruitment, while generated profiles with little or no food-delivery experience led her to add recent DoorDash experience as an inclusion criterion. During interview rehearsal, a simulated response describing decision time when a food preference was already known exposed her question's assumption that users knew what they wanted before opening the app. She therefore revised the question to compare decision time across ordering scenarios. Her follow-up about having no preference elicited frustration with similar-looking options, prompting a new question about app-related frustrations. At the thematic analysis dashboard, the theme about visual cues prompted her to add a probe about which visual elements mattered most. The theme about decision effort and uncertainty also helped her revise the research focus to examine which aspects of DoorDash influenced users' choices and decision time. Across these stages, P4 used system outputs as prompts for reflection and selectively revised her protocol based on her research goals.
 
\subsection{How do users interact with and evaluate \systemName{}? (RQ1)}

\label{sec:rq1}

In this section, we first examine participants’ existing uses of LLMs and how they compared these experiences with using \systemName{}. We then examine how participants interacted with \systemName{}’s features and how they perceived their roles
in supporting protocol reflection and revision. We organize the findings around the benefits and risks participants
perceived while using each feature.

\subsubsection{Participants' Perceived Differences Between \systemName{} and General-Purpose LLMs}

Across the 21 participants, only three had previously used LLMs to simulate participants or personas in their qualitative research. Their prior practices involved either prompting LLMs to generate and role-play personas (P7, P14), or having researchers themselves role-play predefined participant profiles to examine how an AI chatbot responded to users with different socioeconomic and educational backgrounds (P4). These practices relied on manually constructed prompts for specific simulation tasks. In comparison, participants described \systemName{} as bringing these activities into a more structured research workflow that connected literature-grounded population construction, multiple simulated participants, interview rehearsal and thematic review (P6, P8, P12, P15, P21). As P7 described, \systemName{} organized the process ``into phases and in a very clear way,'' reducing the need to manually construct personas, maintain context across prompts, compare responses, and reorganize outputs. At the same time, participants noted that many individual functions could also be reproduced with general-purpose LLMs, which offered greater flexibility for open-ended exploration (P5, P11, P14, P18). Participants also distinguished \systemName{} through its literature grounding and traceability. They valued being able to inspect the sources informing participant construction and trace themes back to individual responses and questions, giving them more context for assessing generated outputs (P3, P7, P9, P14, P16). Overall, participants perceived \systemName{} as a purpose-built workflow that made protocol rehearsal more structured and inspectable.

\subsubsection{Simulated Responses Supported Protocol Reflection but Smoothed Over Interviewing Difficulties}

Participants frequently found \systemName's simulated responses plausible but unsurprising. This finding aligns with prior work showing that simulated responses can resemble human interview accounts closely enough to support sanity checking, while also appearing overly complete, idealized, and less grounded in the depth of lived experience~\cite{hamalainen2023evaluating,kapania2025simulacrum}. Despite their typicality, researchers still found these responses useful for prompting new probes and surfacing gaps in their protocols. At the same time, they identified two important limitations: cooperative responses could mask weak or underspecified questions, while researchers were more likely to dismiss unexpected details when they knew the responses were AI-generated. 

\begin{figure*}[htbp]
    \centering
    \includegraphics[width=0.98\linewidth]{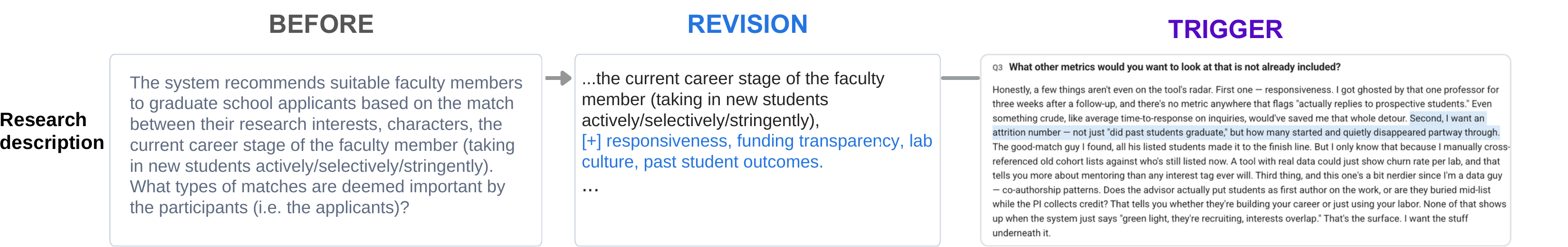}
    \caption{Example of P6's revision. A simulated response surfaced additional criteria for evaluating faculty advisors, prompting him to expand the study's scope to include responsiveness, funding transparency, lab culture, and past student outcomes.}

\Description{The figure presents P6's research description before and after revision, along with the simulated response that prompted the change. His initial description focused on research-interest fit, personality, and faculty career stage. A simulated student identified additional considerations, including responsiveness, co-authorship, lab culture, and past student outcomes. P6 revised the description to incorporate responsiveness, funding transparency, lab culture, and past student outcomes.}
    \label{fig:P6}
\end{figure*}

\paragraph{Benefits: Simulated responses surfaced directions and gaps for follow-up.}
Simulated responses surfaced both additional directions for inquiry and gaps in the existing protocol. P6 initially described his study as examining faculty recommendations based on research-interest fit, personality, and faculty career stage. The simulated students surfaced additional criteria, including “past student outcomes” and “responsiveness and co-authorship patterns.” P6 subsequently expanded the study to also consider past student outcomes, responsiveness and lab culture. Fig. ~\ref{fig:P6} illustrates this revision process. In other cases, researchers identified directions for follow-up by noticing what was absent from the simulation. P9 was studying how users respond to problematic behaviors of autonomous AI agents, with an initial protocol focused on users’ reactions, authority boundaries, and preferred control mechanisms. During rehearsal, she realized that “none of these generated questions or replies mention accountability,” prompting her to add a new question about designer responsibility for an AI agent’s behavior. Across these cases, simulated responses informed protocol revision by surfacing additional lines of inquiry and making missing explanations or perspectives more visible.

\paragraph{Risks: Over-cooperative responses masked question and communication failures.}

At the same time, participants described simulated respondents as unusually articulate, cooperative, and complete, which could make rehearsal less diagnostic. Several noted that AI responses often provided too much information upfront, leaving fewer opportunities to practice probing. P9 explained that the responses often \textit{``covered everything''} she wanted to hear, so she \textit{``hardly wanted to insert any follow-up questions,''} whereas responses in real interviews often emerge \textit{``bit by bit.''} Participants therefore expressed concern that overly complete responses could conceal weaknesses in interview questions or reduce opportunities to practice follow-up strategies (P13, P14, P19).

This smoothness also reduced opportunities to rehearse population-specific communication challenges. P8, who studies Deaf or Hard of Hearing (DHH) participants, noted that \textit{``the AI's linguistic level is too high,''} limiting opportunities to practice the rephrasing, clarification, and conversational repair that may arise in interviews with DHH participants. More broadly, nine of the 21 participants raised concerns that simulations made interviews too easy or failed to preserve realistic interactional difficulty. Several contrasted the smoothness of simulated responses with the messiness of real interviews, where participants may be brief, cautious, incomplete, or require repeated probing (P9, P13, P14, P17, P19). P17 suggested including \textit{``difficult responses or concise responses, unhelpful responses from participants,''} because \textit{``the real world is messy.''}  Together, these findings suggest that useful rehearsal also needs to preserve the uncertainty, incompleteness, and resistance that make follow-up work necessary in actual interviews.

\paragraph{Risks: Synthetic provenance discouraged researchers from pursuing unexpected details}


Participants sometimes noticed clear signs that the responses were AI-generated. When one simulated participant was instructed to be hesitant, P21 observed that \textit{``every single answer starts with an `um,' ''} rather than showing the more uneven variation she expected from human participants. Once participants were reminded that the material was synthetic, this affected how they interpreted and followed up on the responses. P9 noted that she would normally take notes in a real interview, but stopped because “this is an AI-generated conversation” and “taking notes doesn’t have much point.” Knowing the responses were synthetic also changed which unexpected details researchers considered worth pursuing. P17 dismissed a surprising theme during the thematic analysis, explaining, “because it is AI, instead of saying, ‘Let’s explore that further,’ I guess I just said, ‘Oh, it’s AI.’”  Thus, unexpected details that might have prompted further inquiry in a human interview were more easily dismissed when they came from simulated participants.

\subsubsection{Making Populations Legible Broadened Coverage but Could Flatten Human Complexity}

Participants found \systemName's generated subgroups and profiles useful for making abstract target populations more concrete. At the same time, they raised concerns about flattening overlapping identities and reinforcing stereotypical associations, aligning with prior work~\cite{wang2025misportray,cheng2023marked}. Our findings further show that these representations influenced protocol design by surfacing overlooked participant variation and prompting researchers to consider whether simulated diversity could be reflected in actual recruitment.

\begin{figure*}[htbp]
    \centering
    \includegraphics[width=\linewidth]{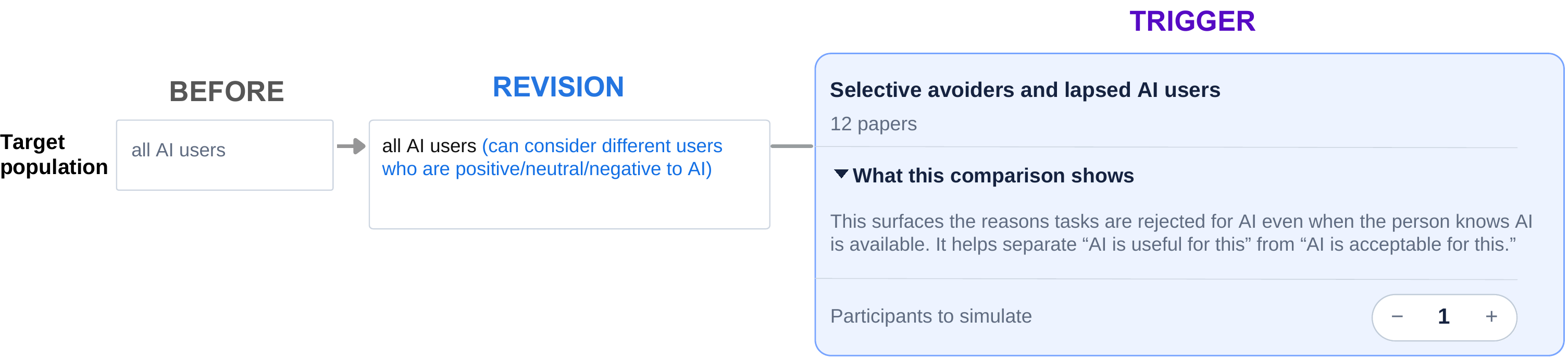}
\caption{Example of P15's revision. The generated subgroup of selective avoiders and lapsed AI users prompted her to expand the target population to include users with positive, neutral, and negative orientations toward AI.}

\Description{The figure presents P15's target population before and after revision, along with the subgroup that prompted the change. Her initial population included all AI users. After reviewing the generated subgroup of selective avoiders and lapsed AI users, she revised the population to consider users with positive, neutral, and negative orientations toward AI.}
    \label{fig:P15}
\end{figure*}

\paragraph{Benefits: Generated subgroups surfaced overlooked participant variation.}
\systemName\ presented researchers with concrete alternatives for reconsidering who their studies should include. P2 was studying how children use GenAI for animated video creation and had initially framed this as an individual activity. After seeing a peer/multi-person subgroup, she noted that \textit{``this peer/multi-person co-creation---I hadn't considered it''} and expanded her study to include group interaction and collaboration, adding a question about whether AI mediated or intensified conflict during collaboration. Generated variation also foregrounded contrastive perspectives. P15 was studying why people use AI for some tasks but not others, with an initial target population of simply all AI users. After seeing an AI-avoider subgroup, she reflected that \textit{``if I were recruiting on my own I'd never have thought to recruit AI avoiders''} and that she could \textit{``look for people from the opposite side.''} She subsequently revised the population to include positive, neutral, and negative orientations toward AI. Fig. ~\ref{fig:P15} illustrates this revision process.

\paragraph{Risks: Simulated diversity exceeded realistic recruitment.}
The range of perspectives available through simulation sometimes exceeded what researchers could realistically recruit. P9 appreciated the international diversity of the generated personas but noted that \textit{``in real collection we can't get anything this international.''} P10 similarly valued the opportunity to consider a clinician perspective relevant to her project, while noting that clinicians were difficult to access and unlikely to be included in her actual recruitment. Simulated diversity therefore broadened the range of perspectives researchers could consider during study design without changing the practical constraints of participant recruitment.

\begin{figure*}[htbp]
    \centering
    \includegraphics[width=\linewidth]{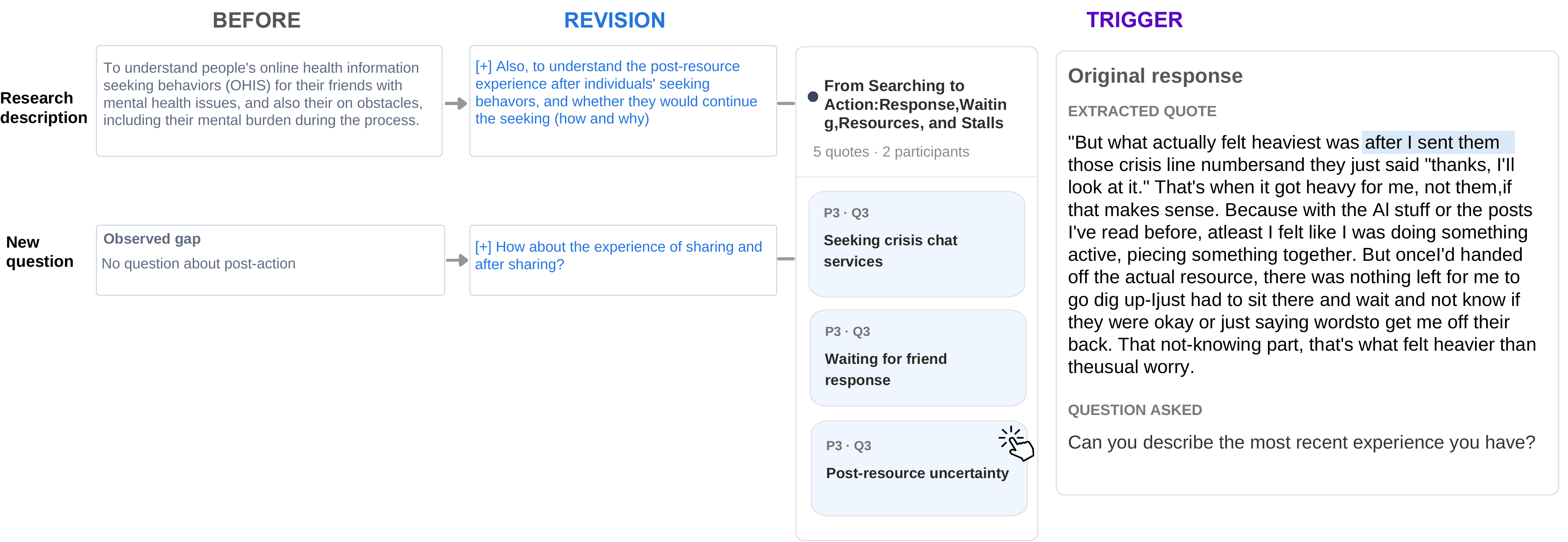}
   \caption{Example of P10's revision. The theme revealed uncertainty and waiting after sharing mental health resources, prompting her to expand the study's scope and add a question about experiences after sharing.}
   \Description{The figure presents three columns: Before, Revision, and Trigger. The original study focused on seeking online health information for friends with mental health concerns but lacked questions about experiences after sharing resources. Revisions, highlighted in blue, extend the research description to include subsequent experiences and continued information seeking, and add a question about sharing and its aftermath. The trigger panel links dashboard codes to a simulated response describing uncertainty while waiting for a friend's reaction after sending crisis resources.}
    \label{fig:P10}
\end{figure*}

\subsubsection{AI-Generated Themes Accelerated Sensemaking but Pre-Structured Interpretation}

Participants used generated themes as provisional shortcuts to identify gaps and consider possible interpretations. However, they also worried that these themes could shape expectations for future human interviews and reduce opportunities to develop their own interpretations.

\paragraph{Benefits: Themes provided provisional shortcuts for identifying gaps and possible interpretations.}
Generated themes gave researchers a fast overview of what their protocols surfaced across responses, making omitted topics and relationships easier to notice. P10's online health-information-seeking protocol initially focused on experiences before and during searching. After seeing a theme about post-results experience, she realized that she had missed an entire stage: \textit{``besides stage one---before searching---and stage two---during searching, I'd add a third stage: after searching.''} She subsequently expanded the research description and added a question about information sharing and what happened after sharing. Fig. ~\ref{fig:P10} illustrates this revision process. At the same time, participants generally treated these themes as provisional cues rather than findings to accept directly. P3 described them as a \textit{``shortcut''} or an \textit{``extremely convenient standby theme''}, while emphasizing the need to remain \textit{``open to other possibility.''}

\paragraph{Risks: Themes could anchor subsequent questioning and interpretation.}
Participants were also concerned that themes could become expectations before any human data had been collected. P2 described this carry-over directly: after using \systemName, \textit{``when I go to observe and interview I may go in carrying some preset themes.''} P10 raised the same concern from the perspective of grounded theory, where researchers should avoid over-presupposing the kinds of answers participants will give. For her, AI-generated themes introduced an additional source of bias that was more difficult to account for than the researcher's own positionality. Participants also worried about confusing synthetic and human material after repeated exposure. P1 described the possibility that researchers might later \textit{``misremember or pay attention to some real answers''} and incorrectly experience a theme as recurring because they had already encountered it in simulated interviews. P21 similarly found the thematic output unsettling because she could already imagine each generated theme appearing in a future results section: \textit{``I could picture each of these as a result in the results section. And that worries me.''}

\subsubsection{Simulation Supported Research Preparation but Required Epistemic Judgment}

\systemName\ provided a preparatory space for researchers to practice interviewing decisions before engaging human participants. Participants described rehearsal as useful for developing adaptive probing and methodological skills. At the same time, they identified epistemic limits to this role: unfamiliar AI-generated suggestions could be difficult to distinguish from meaningful insights, and simulated material could blur the boundary between research preparation and empirical evidence.



\paragraph{Benefits: Rehearsal supported methodological training.}
Participants saw this rehearsal setting as useful for methodological training. P9, a novice qualitative researcher, described learning to identify cues for follow-up probing: \textit{``when I hear certain keywords, I can pose follow-up questions targeted at the keywords in that user's reply.''} P17 explained that she normally plays the participant herself when training students to conduct interviews. With \systemName\ acting as the practice participant, she could instead focus on observing students' interviewing and providing feedback. P3 similarly described \systemName\ as a potential \textit{``qualitative research companion''} given the difficulty of training qualitative researchers. Simulation therefore provided a setting for practicing interviewing decisions and receiving feedback without requiring a human participant for every rehearsal.


\paragraph{Risks: Researchers could not always distinguish unfamiliar insights from AI errors.}
Researchers also faced uncertainty when generated material challenged their existing understanding. P21 asked, \textit{``if something that the AI tells you here challenges your belief, is that because the AI is wrong, or is it because you are wrong? [...] It's actually basically impossible to tell, especially if it's in a new space.''} P17 similarly noted that researchers may be more influenced by generated material when they are unfamiliar with the topic: \textit{``not too sure what the concept is yourself [...] you suddenly read these, and then that's what you expect to find.''} An unfamiliar suggestion could therefore prompt reflection while leaving researchers uncertain whether it revealed a blind spot or an AI error.


\begin{table*}[htbp]
\centering
\footnotesize

\setlength{\tabcolsep}{3.5pt}

\resizebox{\textwidth}{!}{%
\begin{tabular}{l cr cr crrr ccccccc}
\toprule
& \multicolumn{2}{c}{Research context}
& \multicolumn{2}{c}{Target population}
& \multicolumn{4}{c}{Interview questions}
& \multicolumn{7}{c}{Targeted dimension} \\
\cmidrule(lr){2-3}
\cmidrule(lr){4-5}
\cmidrule(lr){6-9}
\cmidrule(lr){10-16}
Participant & Mod. & $\Delta$ words
            & Mod. & $\Delta$ words
            & Mod. & Edited & Added & $\Delta$ words
            & Ctx. & Pop. & Cov. & Cla. & Dep. & Bre. & Rig. \\
\midrule
P1  &            &          &            &          & \checkmark &   & 2 & $+69$     &         &         & \lcheck &         & \lcheck & \lcheck &         \\
P2  & \checkmark & $+4$     & \checkmark & $+5$     & \checkmark & 1 & 2 & $+49$     & \lcheck & \lcheck & \lcheck &         & \lcheck & \lcheck &         \\
P3  &            &          & \checkmark & $+13$    &            &   &   &           &         & \lcheck &         &         &         & \lcheck &         \\
P4  & \checkmark & $+16$    & \checkmark & $+19$    & \checkmark & 2 & 1 & $+23$     & \lcheck & \lcheck & \lcheck &         & \lcheck & \lcheck & \lcheck \\
P5  &            &          &            &          &            &   &   &           &         &         &         &         &         &         &         \\
P6  & \checkmark & $+8$     &            &          & \checkmark & 2 &   & $+4$      & \lcheck &         & \lcheck & \lcheck & \lcheck &         &         \\
P7  & \checkmark & $+5/-1$  & \checkmark & $+3/-2$  & \checkmark &   & 1 & $+16$     &         & \lcheck & \lcheck &         &         & \lcheck &         \\
P8  & \checkmark & $+1$     & \checkmark & $+10$    &            &   &   &           & \lcheck & \lcheck &         &         &         &         & \lcheck \\
P9  &            &          &            &          & \checkmark &   & 1 & $+38$     &         &         & \lcheck &         & \lcheck & \lcheck &         \\
P10 & \checkmark & $+20$    &            &          & \checkmark & 1 & 1 & $+37/-14$ & \lcheck &         & \lcheck &         &         & \lcheck & \lcheck \\
P11 &            &          &            &          &            &   &   &           &         &         &         &         &         &         &         \\
P12 & \checkmark & $+51$    &            &          & \checkmark &   & 2 & $+30$     & \lcheck &         & \lcheck &         & \lcheck &         &         \\
P13 &            &          & \checkmark & $+4$     &            &   &   &           &         & \lcheck &         &         &         &         &         \\
P14 & \checkmark & $+39$    &            &          & \checkmark & 2 &   & $+9$      & \lcheck &         &         & \lcheck &         &         & \lcheck \\
P15 &            &          & \checkmark & $+9$     & \checkmark & 3 &   & $+17/-1$  &         & \lcheck & \lcheck &         & \lcheck & \lcheck & \lcheck \\
P16 & \checkmark & $+11$    & \checkmark & $+13/-4$ &            &   &   &           & \lcheck & \lcheck &         &         &         & \lcheck & \lcheck \\
P17 &            &          &            &          &            &   &   &           &         &         &         &         &         &         &         \\
P18 & \checkmark & $+8$     & \checkmark & $+15$    &            &   &   &           & \lcheck & \lcheck &         &         &         & \lcheck & \lcheck \\
P19 &            &          & \checkmark & $+17$    &            &   &   &           &         & \lcheck &         &         &         & \lcheck &         \\
P20 &            &          &            &          &            &   &   &           &         &         &         &         &         &         &         \\
P21 & \checkmark & $+19$    & \checkmark & $+8$     & \checkmark &   & 2 & $+38$     & \lcheck & \lcheck &         &         & \lcheck & \lcheck & \lcheck \\
\midrule
Participants ($n$) & 11 & & 11 & & 11 & 11 & 12 & & 10 & 11 & 9 & 2 & 8 & 12 & 8 \\
\bottomrule
\end{tabular}%
}
\caption{Participants' revisions to the initial protocol. On the left, ``Mod.'' indicates whether a protocol component was modified, and $\Delta$~words reports the number of words added ($+$) and removed ($-$) relative to the original. For interview questions, ``Edited'' counts existing questions whose wording was changed, while ``Added'' counts newly inserted questions. On the right, checkmarks indicate which protocol-quality dimensions were targeted by each participant’s revisions. The bottom row reports the number of participants with at least one coded revision in each dimension: Ctx.\ = research context, referring to clarity about the settings and situations within the study's scope; Pop.\ = target population, referring to clarity about who is included in the study; Cov.\ = coverage, referring to whether key aspects of the research question are addressed; Cla.\ = clarity, referring to the wording and comprehensibility of interview questions; Dep.\ = depth, referring to the extent to which questions support rich and detailed responses; Bre.\ = breadth, referring to the range of perspectives or experiences captured; and Rig.\ = rigor, referring to the methodological coherence of the protocol. A checkmark indicates that at least one coded revision targeted the corresponding dimension. No revision was coded as related to feasibility, so that column is omitted. These codes indicate the dimensions related to revisions, not whether the revisions improved protocol quality.}
\label{tab:participant-revisions}

\end{table*}

\subsection{How does rehearsal with \systemName{} shape users’ reflections and protocol revisions? (RQ2)}

\label{sec:rq2}

In this section, we first examine revisions in participants' protocols, including which parts of the protocol they changed and which dimensions of protocol quality these revisions affected. We then examine participants' perceived changes in protocol quality and revision readiness based on their pre and post self-assessment surveys.

\begin{figure*}[htbp]
    \centering
    \includegraphics[width=\linewidth]{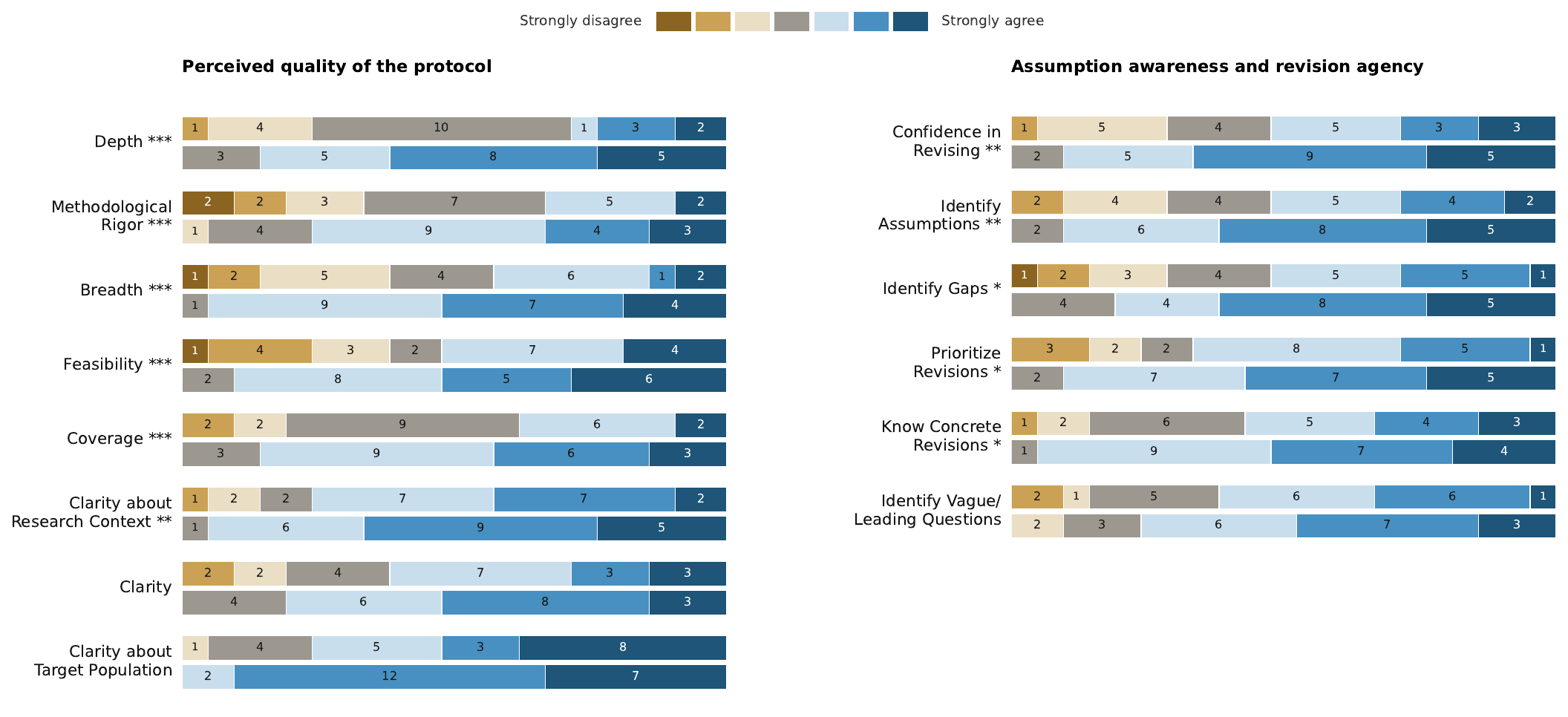}
   \caption{Distribution of participants' ratings of their research protocols before (top of each item) and after (bottom of each item) using \systemName. Participants reported improvements in clarity about the research context, coverage, depth, breadth, methodological rigor, and feasibility, as well as their ability to identify assumptions and gaps, identify concrete revisions, prioritize revisions, and revise with confidence. (*: $p<.05$, **: $p<.01$, ***: $p<.001$).}
   \Description{Two side-by-side groups of horizontal stacked bar charts compare participants' ratings before and after using \systemName\ across 14 seven-point Likert-scale measures. The left panel presents perceived protocol quality, including research-context clarity, target-population clarity, coverage, question clarity, depth, breadth, methodological rigor, and feasibility. The right panel presents assumption awareness and revision agency, including identifying assumptions, identifying gaps, identifying vague or leading questions, knowing concrete revisions, prioritizing revisions, and confidence in revising. For each measure, the top bar shows ratings before \systemName\ and the bottom bar shows ratings after \systemName\. Ratings shift toward stronger agreement after use, with significant increases in research-context clarity, coverage, depth, breadth, methodological rigor, feasibility, identifying assumptions and gaps, knowing and prioritizing concrete revisions, and confidence in revising.}
    \label{fig:rq1}
\end{figure*}

\subsubsection{Protocol Revisions Spanned Multiple Dimensions.}

We first examined how participants revised their protocols, as summarized in Table~\ref{tab:participant-revisions}. Seventeen of the 21 participants modified at least one of the three protocol components: research context, target population, or interview questions. Among these 17 participants, they modified an average of 1.94 components ($SD=0.75$) and changed 38.24 words across the three components ($SD=24.65$). Research context and target population were each revised by 11 participants, with an average of 10.76 ($SD=14.73$) and 7.18 ($SD=6.93$) changed words per reviser, respectively. Interview questions were also revised by 11 participants and showed the greatest amount of textual change ($M=20.29$, $SD=22.02$ words). Across participants, 11 existing questions were edited and 12 new questions were added. Four participants left their protocols unchanged for different reasons. P5 and P20 sought feedback from simulated users to improve systems they had designed, but felt that the simulation lacked sufficient context to inform protocol revisions. P17 considered exploring an unexpected response but chose not to revise because it came from an AI simulation. P11, an expert in the field, considered his initial interview protocol already well developed and saw little room for revision.

To characterize the dimensions represented in these changes, two researchers independently coded each revision move using the protocol-quality dimensions measured in our survey. Coding was based on both the revised protocol artifacts and participants' think-aloud explanations of their revisions. Each revision could be related to several dimensions. The two researchers discussed disagreements and reached consensus on the final codes. The coded revisions spanned multiple dimensions of protocol design. Breadth was represented in revisions by the largest number of participants ($12$), followed by target population ($11$), research context ($10$), coverage ($9$), depth ($8$), and methodological rigor ($8$). In terms of coded revision moves, depth had the largest number of moves ($+15$), followed by target population and breadth ($+14$ each), coverage ($+11$), research context and methodological rigor ($+10$ each), and clarity ($+3$). These results show that participants' revisions extended beyond question wording to include who should be studied, what the protocol should cover, and the breadth and depth of responses it should support.

\subsubsection{Revisions Took Several Recurring Forms.}
Beyond identifying which dimensions were involved in participants' revisions, we examined how those changes were made. Across dimensions, the revisions clustered into three recurring patterns: redrawing study and population boundaries, expanding what the protocol could surface, and improving alignment across protocol components.

\paragraph{Redrawing study and population boundaries.}
Participants revised both what counted as part of the study and who counted as a relevant participant. For research context, they broadened or narrowed the scope by adding or constraining settings, phases, constructs, or situations. Target-population revisions similarly changed inclusion boundaries by adding contextual requirements, differentiating subgroups, reframing population definitions, or adjusting sample composition. These changes often replaced coarse population descriptions with distinctions based on role, prior experience, behavior, attitude, or research context. For example, P4 added recent DoorDash use as an inclusion criterion and differentiated participants by price and time sensitivity, while P16 replaced an age-based boundary with distinctions based on workplace type and design domain.

\paragraph{Expanding what the protocol could surface.}
Participants expanded their protocols in three ways. Coverage revisions introduced topics, stages, or constructs that had previously been absent, such as trust, responsibility, friction, or group conflict. Depth revisions elaborated existing lines of inquiry through probes asking participants to explain reasons, describe processes, clarify meanings, or recount concrete experiences. Breadth revisions introduced variation across people, situations, and viewpoints, such as contrasts between novice and experienced participants, users and AI avoiders, or individual contributors and managers. For example, P9, who was studying users’ reactions to autonomous AI agents in academic collaboration scenarios, added a question about designers’ responsibility for agents’ behavior and a probe asking participants to clarify key terms in their responses. P12, who was studying how AI processing wait times shape creative ideation, added a question asking participants to walk through a specific long-wait episode to elicit concrete details about how waiting unfolded within their workflow.

\paragraph{Improving alignment across protocol components.}
Other revisions focused on internal coherence across the protocol. Clarity-related revisions made question referents more specific or opened closed questions to elicit explanation. Rigor-related revisions aligned questions with research objectives, population criteria with study scope, or question wording with a particular participant group. For example, P14 was studying how learners perceive AI-powered educational tools. She revised broad questions about the tool to focus specifically on its design and layout, bringing the interview questions into closer alignment with her design-focused study objective.

\subsubsection{Participants’ Perceived Improvements Only Partially Aligned with Documented Revisions}

We also examined participants' self-reported perceptions of protocol quality and revision readiness before and after using \systemName. The results are shown in Fig. ~\ref{fig:rq1}. These measures capture participants' self-perceived improvements but do not necessarily indicate actual improvements in protocol quality or revision ability.

After using \systemName, participants rated their protocols more positively in terms of coverage, depth, breadth, methodological rigor, clarity about the study setting, and feasibility. They also reported greater awareness of assumptions and contextual gaps, a clearer sense of concrete revisions and priorities, and greater confidence in revising. Changes were less evident for perceived question clarity, clarity about the target population, and the ability to identify vague, leading, or incomplete questions. We also examined correlations between participants' backgrounds and changes in their self-reported ratings of protocol quality and revision readiness but found no statistically significant associations.

We then examined whether  participants' own perceptions of change aligned with the researchers' coded revisions. We compared whether participants perceived a change in each protocol-quality dimension with whether our coding identified a revision in that dimension. We assessed agreement using Cohen's $\kappa$. Overall, the two sources showed limited alignment: 51.8\% of participant–dimension pairs agreed ($\kappa = 0.14$). Most disagreements occurred when participants' ratings indicated changes in a dimension without a corresponding coded revision (70/168, 41.7\%), whereas coded revisions without corresponding changes in self-assessments were less common (11/168, 6.5\%). At the participant level, however, the number of dimensions with self-reported change was positively correlated with the number of dimensions represented in coded revisions ($\rho = 0.58$, $p < 0.01$). Participants who revised more dimensions therefore also tended to perceive broader changes in their protocol quality, even though the specific dimensions they reported did not always match those identified in the coding.

\begin{figure*}[t!]
    \centering
    \includegraphics[width=\linewidth]{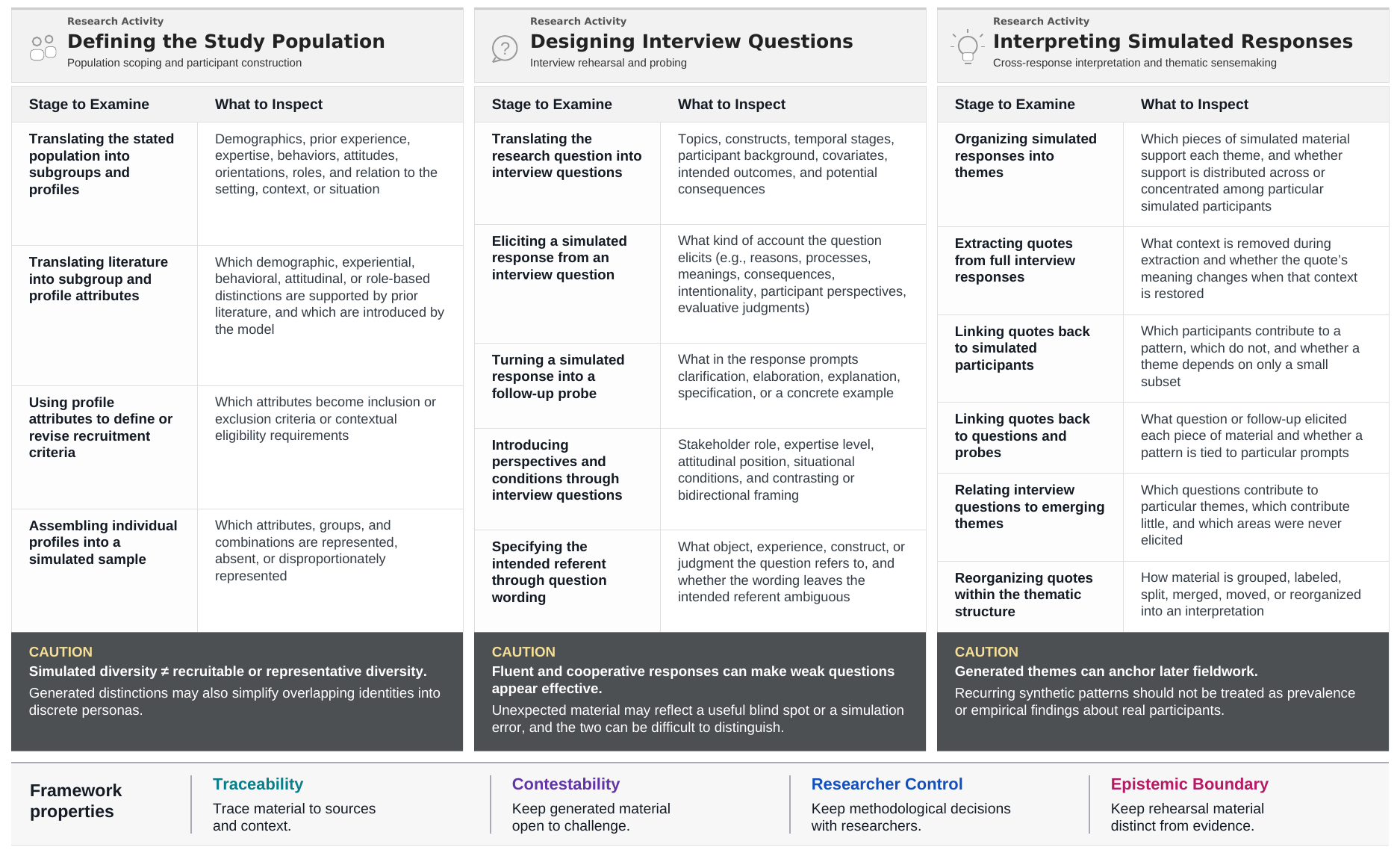}
    \caption{An \textit{Inspectable Rehearsal Framework} for qualitative interview design. The framework organizes three activities through which LLM-generated materials can inform early-stage research design: defining whom to study, designing what to ask, and examining what the protocol elicits. Within each activity, it identifies concrete relationships to inspect and highlights cautions that arise when simulated materials inform methodological decisions.}
    \Description{The figure presents three research-activity columns. Defining the Study Population covers constructing subgroups and profiles, grounding attributes in literature, revising recruitment criteria, and assembling a simulated sample. Designing Interview Questions covers translating research questions into interview questions, examining what responses questions elicit, developing follow-up probes, introducing perspectives and conditions, and clarifying question wording. Interpreting Simulated Responses covers constructing themes, examining quote context, tracing quotes to participants and prompts, identifying which questions contribute to themes, and reorganizing material. Caution panels beneath the columns address simplified identities and limits of simulated diversity; cooperative responses that conceal weak questions and unexpected material that is difficult to evaluate; and thematic anchoring and the risk of treating recurring synthetic patterns as empirical evidence. Four cross-cutting framework properties are shown beneath the activities: traceability, contestability, researcher control, and epistemic boundary.}
    \label{fig:framework}
\end{figure*}

\section{Discussion} 

\subsection{\textit{Inspectable} Rehearsal Framework for Qualitative Interview Design}

Our findings suggest that simulated participants can support methodological reflection even when their fidelity is limited. Participants readily noticed that \systemName's responses were often unusually articulate and cooperative and that they did not fully capture some group-specific experiences. Yet participants still used these materials to reconsider whom to recruit, what to ask, and what their questions might elicit.

We therefore propose the concept of \textit{inspectable material}: provisional artifacts whose content, provenance, and relationships can be examined and critically assessed as researchers make methodological decisions. This concept builds on prior definitions of inspectability, which emphasize making connections between source material and interpretation traceable and exposing where analytic judgment enters the process~\cite{vermeeren2008usability}. Participants' comparisons with general-purpose LLMs further support this framing. They valued \systemName{} as a purpose-built workflow that organized familiar LLM capabilities into stages where sources, intermediate materials, and downstream interpretations could be inspected. Rather than treating generated artifacts as conclusions to accept, we argue that they should remain inspectable so that researchers can examine their provenance, relationships, and methodological implications.

Based on our coding of participants' revised protocols and observations of their interactions with the workflow, we derived a set of concrete aspects that should remain inspectable during research rehearsal. We organize these aspects into an \textit{Inspectable Rehearsal Framework} for LLM-supported qualitative interview design, shown in Fig.~\ref{fig:framework}. The framework spans three research activities: defining the study population, designing interview questions, and interpreting simulated responses.

When defining the study population, researchers examined population attributes such as age, prior experience, behavior, attitudes, professional roles, relationships to a setting, and contextual eligibility. These distinctions affected who was included, how participants were differentiated, and how the intended sample was composed. When designing interview questions, researchers examined what a question targeted, what kind of account it invited---such as a reason, process, meaning, consequence, or judgment---whose perspective it foregrounded, what conditions it assumed, and what its wording referred to. When considering \textit{what the protocol elicited}, inspection extended beyond individual responses to how interpretations were constructed: which participant and question produced a quote, what context surrounded it, which materials contributed to a theme, and how those materials were grouped. Fig.~\ref{fig:framework} organizes these as concrete points of inspection across population construction, interview rehearsal, and thematic sensemaking.

These elements need to remain inspectable because synthetic material can influence methodological decisions while also being easy to over-interpret. Generated population distinctions can broaden attention to diversity while flattening overlapping identities or suggesting combinations that may be difficult to recruit in practice. Simulated responses can reveal missing directions or opportunities for probing, but their fluency and cooperativeness can also make vague or weak questions appear effective. Unexpected responses create a different ambiguity: researchers may treat them as prompts for further inquiry, dismiss them as model errors, or over-interpret them as information about real participants. At the level of thematic analysis, aggregation can make patterns easier to notice while also making provisional synthetic regularities appear more stable than they are. Recurrence across simulations therefore does not establish prevalence, and generated themes do not constitute empirical findings about the target population.

Across these stages, our findings point to four properties for supporting inspection. First, \textit{traceability} allows researchers to follow generated material back to its sources and context, from literature to profile attributes, from questions to responses and probes, and from themes to quotes, questions, and participants. Second, \textit{contestability} keeps generated distinctions and interpretations open to challenge: researchers should be able to reject, revise, regroup, or replace what the system proposes. Third, \textit{researcher control} preserves methodological authority over population definitions, follow-up decisions, protocol revisions, and thematic interpretation. Inspectability should not merely expose how an output was produced while leaving the resulting methodological decision fixed. Finally, an \textit{epistemic boundary} keeps rehearsal material distinct from empirical evidence. Synthetic profiles, responses, and recurring themes can surface assumptions worth examining, but they cannot establish who real participants are, how they would respond, or what patterns would emerge in fieldwork.

This framing shifts the design goal for simulated-participant systems. Rather than focusing only on realism, rehearsal-oriented systems should make the methodological consequences of simulation inspectable: what population distinctions are introduced, what aspects of a question are exercised, what kinds of responses are elicited, and how interpretations are constructed. In this role, simulation functions as an epistemic scaffold for preparing qualitative research rather than as a substitute for empirical fieldwork. Its value lies in making researchers' assumptions, omissions, and methodological choices available for examination before data collection.

\subsection{Design Implications}

\subsubsection{Designing for Epistemic Friction}

Our findings show that participants wanted simulated interviews to preserve more of the uncertainty and difficulty of interviewing real people. Qualitative interviewing is inherently interactional: extensive knowledge of a research topic does not guarantee that participants will understand questions or respond as researchers intend~\cite{yeong2018interview}. Questions that appear clear to researchers may still be interpreted differently, requiring clarification, probing, and adaptation during the interview~\cite{dejonckheere2019semistructured,knott2022interviews,conrad2000clarifying}. Such differences also reflect participants' cultural and linguistic backgrounds, educational experiences, familiarity with terminology, and lived contexts~\cite{yeong2018interview,squires2008crosslanguage}. Misunderstandings and incomplete answers can therefore expose gaps between researchers' assumptions and participants' understandings. By readily inferring intended meanings and providing complete answers, overly cooperative simulations can conceal these gaps and limit opportunities to practice conversational repair.

Simulated-participant systems should therefore support \textit{epistemic friction}: resistance that brings researchers' assumptions into question and prompts reconsideration. Drawing on Medina's account of epistemic friction through engagement with differing perspectives~\cite{medina2013epistemology}, we apply this concept to protocol rehearsal, where unexpected interpretations or challenges to a question's premise can reveal limitations in how researchers frame their inquiry. Systems could deliberately construct participants who give brief or vague answers, misunderstand terminology, or interpret questions differently. They could also introduce unfamiliar perspectives or conflicting values that challenge assumptions embedded in the protocol. During rehearsal, these responses would require researchers to decide when to clarify, probe, rephrase, or redirect the conversation. The epistemic value of these encounters lies in what they reveal about a protocol's assumptions and the judgments they require researchers to make. This direction builds on work on cognitive forcing functions that encourages deliberate engagement with AI outputs~\cite{bucinca2021trust}, extending its focus to examining and adapting research questions. Researchers could practice recognizing a mismatch, investigating its source, and revising their wording or approach. These generated challenges provide possibilities for reflection whose relevance still requires assessment with human participants.

\subsubsection{Supporting Critical Engagement Through Researcher Control}
Our findings show that even though participants found AI-generated material useful, some participants remained reluctant to adopt or seriously consider these ideas because they were AI-generated. This tension aligns with research on algorithm aversion and AI disclosure effects, which suggests that people may discount otherwise useful outputs because of their AI provenance~\cite{castelo2019task,lim2024source,meincke2026advice}. In research practice, such skepticism serves a methodological purpose: researchers must scrutinize generated material for unsupported assumptions and assess its relevance to their study contexts. Design should therefore support critical engagement without eliminating this caution.

Prior work shows that greater control over algorithmic outputs can reduce resistance and increase willingness to engage with recommendations~\cite{cheng2023overcoming,westphal2023decision,fink2024let,schaap2024abc}. Building on this work, AI-supported research systems should support selective revision and incorporation of generated suggestions. Researchers could edit proposed questions, remove embedded assumptions, or integrate relevant elements into existing protocols. Rather than requiring researchers to accept or reject an AI suggestion as a whole, such controls allow generated material to remain subordinate to researchers' methodological judgment.
 
Critical engagement also extends to how researchers assess their own protocols. Although participants who perceived changes across more protocol-quality criteria tended to revise more criteria, their perceptions of which criteria had changed differed from those represented in their documented revisions. This echoes prior concerns about discrepancies between AI-assisted performance and self-assessment~\cite{fernandes2025ai}, as well as the importance of calibrating confidence in one's own judgments~\cite{ma2024you}. In qualitative interview rehearsal, protocol quality cannot be readily evaluated against a correct answer, making it difficult to distinguish whether perceived improvements reflect a deeper understanding of existing protocol choices or unwarranted confidence in them. Rehearsal systems should therefore encourage researchers to critically examine the basis of their judgments, rather than treating perceived improvements as evidence of better protocol quality. For example, systems could help researchers compare their perceived improvements with documented revisions, revisit the simulated materials that informed their assessments, and reflect on why they chose to revise or retain particular questions.

\subsubsection{Designing AI-Supported Research Workflows as Inspectable Orchestration}

Our findings suggest that the value of an integrated AI-supported research workflow lies in reducing the coordination required when working with general-purpose LLMs. \systemName\ connected activities such as literature exploration, participant construction, interview rehearsal, and thematic inspection while preserving context across stages. This reduced the need for researchers to repeatedly reconstruct prompts and, at the same time, externalized intermediate decisions about research goals, target populations, questions, and interpretations.

However, integration also creates dependencies between stages. In \systemName, early decisions about the research framing, literature, or participant subgroups could influence profiles, simulated responses, and subsequent analysis. Participants therefore raised concerns that assumptions introduced upstream could propagate through later outputs and become harder to recognize as they were repeatedly reproduced. Future AI-supported research systems should make these dependencies inspectable by allowing researchers to trace which inputs, sources, and prior outputs contributed to a generated artifact, revise upstream decisions, and see how those changes affect downstream results. Such provenance should also extend to data handling, making clear what research information is stored, reused, or passed between system components. Rather than optimizing only for seamless automation, research workflows should preserve researchers' ability to inspect and intervene in how information and assumptions move through the system.

\subsubsection{Supporting Interactional Rehearsal Through Multimodal Simulation}

Our findings show how text-based simulation can flatten interactional states. One participant noted that a ``hesitant'' simulated participant began every answer with ``um.'' This reduced hesitation to a repetitive linguistic cue rather than an interactional moment that required interviewer judgment. In qualitative interviewing, silence is meaningful and can reflect resistance, reflection, or shifts in interviewer--interviewee relations~\cite{bengtsson2018analysing}. Research on conversational systems similarly shows that turn-taking and response timing shape how conversational partners coordinate interaction~\cite{skantze2021turntaking,jiang2026hear}. These findings suggest that future interview-rehearsal systems should represent hesitation through temporal and paralinguistic variation, including pauses, shifts in speaking pace and intonation, intentional or unintentional topic changes, and delayed continuation, rather than repeated filler words.

Interview rehearsal should also extend beyond text to support multimodal interaction. Incorporating spoken interaction could better reproduce the temporal and interactional demands of live interviews. Interview-rehearsal systems could allow researchers to move between text and voice depending on what they want to practice. Text can foreground question wording and response content, whereas voice can foreground delivery, pacing, and turn management. Transcripts and audio replay could further support reflection on these interactions.

\section{Limitations and Future Work}

While our study demonstrates the potential of \systemName\ for supporting researchers in rehearsing and refining qualitative interview protocols, several limitations are noteworthy. First, we designed and evaluated \systemName\ specifically in the context of interview-based qualitative research. We chose interviews as an initial setting because protocol development requires researchers to anticipate how questions may be interpreted, where follow-up may be needed, and what perspectives may be overlooked. However, the extent to which our findings generalize to other qualitative methods remains to be further examined. Future work could investigate how simulation-based rehearsal may support methods such as focus groups, contextual inquiry, think-aloud studies, or usability studies, where researchers must prepare not only questions but also participant interactions, tasks, and study situations.

Second, the current version of \systemName\ primarily relies on textual descriptions of the study context and supports text-based simulated interactions. This representation provides limited support for system-oriented qualitative studies in which participants' responses are shaped by their interactions with interfaces, prototypes, or other research artifacts. Future iterations of \systemName\ could incorporate multimodal study materials, such as interface screenshots, interactive prototypes, task flows, or screen recordings, allowing simulated participants to respond to and interact with these materials during rehearsal. Such extensions would enable researchers to rehearse not only how participants might respond to interview questions, but also how qualitative inquiry may unfold around situated interactions with a system.

Third, our pre–post evaluation did not include a comparison condition, so the observed changes in perceived protocol quality and revision readiness cannot be attributed uniquely to Rehearsim. They may also reflect repeated reflection and the act of revising the protocol. Fourth, our background analyses were exploratory and conducted with a small sample across multiple measures; these associations should therefore be treated as hypothesis-generating and require replication with larger samples. 
\section{Acknowledgments}

The authors used GPT-6 to assist in creating illustrative storyboards. These visuals are conceptual only and were not used as research data or evidence.

\bibliographystyle{ACM-Reference-Format}
\bibliography{sample-base}

\newpage

\appendix

\section{Appendix}
\subsection{Rehearsim Prompt Contracts and Runtime Details}
\label{app:system-details}

We organize these materials by the three system modules because each module passes a persistent artifact to the next. Complete system prompt bodies are provided in the accompanying \emph{Rehearsim Prompts} supplementary material; the descriptions below document each contract's runtime inputs, outputs, and application-level validation.

\subsection{Shared Prompt Conventions and Contract Map}
\label{app:prompt-map}

Rehearsim implements model-facing operations as versioned prompt contracts. Each contract states a role, task, structured context, constraints, output format, and final self-check. Values in the context are serialized as data, and structured outputs are constrained by provider-side JSON schemas. The application records the prompt version and model with every generated artifact, allowing a later stage to verify that its inputs were produced under the versions frozen for the current iteration.

Table~\ref{tab:prompt-map} maps the contracts used by the interface to their function and prompt basis. A prompt basis identifies the published task structure adapted by Rehearsim; empirical claims from the source paper remain scoped to its original study. C02, C05a, and C05c are Rehearsim-authored contracts because their tasks concern the system's own comparative sampling workflow.

\begin{table*}[ht]
  \centering
  \scriptsize
  \setlength{\tabcolsep}{4pt}
  \renewcommand{\arraystretch}{1.12}
  \caption{Rehearsim's active prompt contracts, grouped by system module. ``Basis'' names the published prompt family whose task structure was adapted; Rehearsim-authored contracts are identified explicitly.}
  \label{tab:prompt-map}
  \begin{tabularx}{\textwidth}{@{}p{0.9cm}p{3.2cm}X p{3.4cm}@{}}
    \toprule
    \textbf{ID} & \textbf{Operation} & \textbf{Structured input and output} & \textbf{Basis} \\
    \midrule
    C01 & Protocol extraction & Uploaded document $\rightarrow$ participant-facing questions in document order & MatViX information extraction \\
    C02 & Sampling-frame derivation & Study description and target population $\rightarrow$ dimensions, subgroups, screeners, and comparison rationales & Rehearsim-authored and lab-iterated \\
    C03 & Subgroup evidence selection & Research question, subgroup, aspect, and retrieved snippets $\rightarrow$ source-bounded findings and copied excerpts & Unstructured Evidence Attribution \\
    C05a & Roster variation planning & Requested subgroups and settled participants $\rightarrow$ one admissible variation direction per subgroup & Rehearsim-authored \\
    C05c & Participant envelope & One subgroup, its screeners, evidence, and variation direction $\rightarrow$ participant attributes with a basis per attribute & Rehearsim-authored \\
    C05b & Persona crafting & Warranted attributes, subgroup evidence, and study stimulus $\rightarrow$ participant profile & SPASM persona crafting \\
    C06 & Prepared background & Participant profile and assigned experienced claims $\rightarrow$ first-person background used during the interview & SPeCtrum self-introduction \\
    C07 & Participant answer & Profile, background, stimulus, claim relations, prior turns, and current question $\rightarrow$ one answer & ThinkPersona, SYCON-Bench, SPASM, and SPeCtrum \\
    C08--C09 & Follow-up and response check & Visible exchange $\rightarrow$ an optional probe; generator context plus response $\rightarrow$ checklist judgments & Rehearsim follow-up taxonomy and CheckEval \\
    C10--C12 & Quote and code extraction & Included transcripts $\rightarrow$ exact quote notes, inductive labels, and fixed-code assignments & Unstructured Evidence Attribution, HICode, and SCALE \\
    C13--C15 & Theme construction & Existing labels, quote notes, and preserved researcher placements $\rightarrow$ themes and an unassigned set & HICode clustering and reassignment \\
    C16 & By-question synthesis & One question and its included answers $\rightarrow$ a source-bounded summary & Unstructured Evidence Attribution \\
    C17 & Prompt-quality review & Study description, population, stage, and interview questions $\rightarrow$ prompt-quality observations & SCULPT preliminary prompt critique \\
    \bottomrule
  \end{tabularx}
\end{table*}

\subsection{Comparative Sampling and Participant Construction}
\label{app:participant-contracts}

C02 output passes through application-level invariants before it becomes a sampling frame. The supplied target population is reinserted when the model omits the primary anchor, and normalization retains exactly one primary group and at most three priority groups. Every group begins with a participant count of zero, leaving roster allocation to the researcher.

C03 returns one finding and supporting excerpt for each selected paper. After parsing, Rehearsim verifies that the excerpt occurs in the snippet supplied to the model. A failed match removes the excerpt while retaining the summary without quoted support. The graph separately records an aspect whose search completed without evidence and leaves an unattempted aspect unset. This distinction allows partial retrieval to resume without making an unfinished search appear complete. Researchers can also exclude a paper from grounding; this selection enters the profile-source hash and invalidates profiles built from an earlier evidence set.

The participant-construction prompts divide information by decision scope. C05a sees the complete requested roster and settled profiles, but receives no retrieved evidence; it assigns regions within subgroup screeners. C05c sees one subgroup's screeners, evidence, and assigned region, then drafts all members of that subgroup together. For age, gender, occupation, and location, application code stores either a basis copied from supplied text or an explicit plausible-completion label. A distinguishing detail requires a supplied basis, while interview-manner attributes carry a generation rationale. C05b receives the surviving fields and subgroup findings to craft the interview-facing profile. Participant model families are assigned with a subgroup-offset rotation and remain fixed through profile, background, and interview generation.

\paragraph{Subgroup frame details}
C02 places subgroups on two axes. Core subgroups vary within the named population, and peripheral subgroups cover adjacent roles the researcher did not examine. When a population consists several roles, such as children and their teachers, only the role addressed by the research question is anchored while the other roles become subgroups of their own. The queries built for C03 are deliberately short, which only contains a subgroup name plus one situating clause, becauselonger queries dilute the relevance ranking of the snippet service, and the interview questions are never part of a query.

\paragraph{Profile details}
C05c may add one distinguishing detail per participant. The detail must derive from a specific finding and is otherwise omitted, and no participant is written as the average or the extreme of its subgroup. Response-style attributes are drawn from a curated set of coherent combinations and assigned by a seeded permutation over the roster, so every combination appears once before any repeats, and the coverage is logged. C06 runs on the participant's assigned model, receives only the findings that the seeded hash marked as personally experienced, and must cover each of them, so its length follows coverage rather than a word budget. Citation information is removed before generation, so the background never reports what research shows.

\subsection{Interview Preparation and Rehearsal}
\label{app:interview-contracts}

The prepare snapshot records the study description, population, stimulus, ordered questions, selected paper identifiers, participant roster, model assignments, generation seed, fifteen prompt versions, and the participant--claim relation version. Before claiming a run job, Rehearsim checks this snapshot against the current roster and the prompt versions stored on every participant background. A mismatch stops the run before any new transcript is written.

Subgroup findings receive digest-derived claim identifiers. A seeded hash over the relation version, iteration seed, participant identifier, and claim identifier assigns each pair as experienced, context only, unknown, or out of scope. C06 voices only experienced claims in the participant's prepared background. C07 may also receive context-only claims as surrounding knowledge, while unknown and out-of-scope claims are filtered before prompt construction. Identical inputs reproduce the same assignment.

Answer constraints.The participant may not invent events, relationships, or institutions. A session that every participant attended counts as a fact, while specific events
inside it does not. The participant's relationship to each stimulus claim must be paraphrased in
their own words and never reported as a research finding where several readings are supported, the less obvious one is preferred, and repeating earlier phrasing is disallowed.

Table~\ref{tab:interview-boundaries} summarizes the information boundary for each operation in an interview chain. Of particular importance, C08 receives the visible exchange without the hidden profile or literature context, while C09 receives the same participant evidence available to C07. The follow-up decision is therefore grounded in words the researcher can also see, and the response check evaluates the answer against the context that generated it.

\begin{table*}[t]
  \centering
  \small
  \setlength{\tabcolsep}{5pt}
  \renewcommand{\arraystretch}{1.12}
  \caption{Information supplied to each operation in an interview trajectory.}
  \label{tab:interview-boundaries}
  \begin{tabularx}{\textwidth}{@{}p{3.0cm}X X@{}}
    \toprule
    \textbf{Operation} & \textbf{Context supplied} & \textbf{Result used downstream} \\
    \midrule
    Participant answer (C07) & Approved profile, prepared background, study stimulus, participant--claim relations, prior visible turns, and current question & Main response appended to the participant's ordered history \\
    Follow-up decision (C08) & Current question, latest answer, and prior visible interview turns & Zero or one neutral probe attached to the main question \\
    Response checklist (C09) & The participant evidence supplied to C07, conversation history, current question, and response & Independent checklist fields shown during response review \\
    Researcher follow-up & The participant's stored context and interview history through the selected question & Researcher-authored probe and answer appended to the same trajectory \\
    \bottomrule
  \end{tabularx}
\end{table*}

A researcher-authored follow-up uses the participant's stored context and accumulated history through the selected question. Its answer passes through the same response check as a scripted turn and is appended to the existing trajectory. This transcript mutation invalidates the derived quote notes, codes, themes, and prompt feedback so the next analysis pass reflects the added exchange.

\subsection{Cross-Response Sensemaking and Revision}
\label{app:analysis-contracts}

C10 processes transcript responses in participant-scoped batches and returns at most three passages per response. Application code verifies each passage as a substring of its source response and discards a failed match. In a parallel path, C11 generates short inductive labels from transcript segments, while C12 constrains each fixed-code assignment to an enum built from the current codebook. Quote extraction and coding therefore retain separate identifiers and failure states.

C13 groups labels within each participant and contributes only their order to the global pass. C14 consolidates the complete ordered label set into at most eight themes, accepting each input label at most once. C15 receives the resulting theme codebook and assigns the remaining quotes one at a time. An output outside that codebook is rejected, and the corresponding quote enters \emph{Uncertain} through a set difference over unassigned identifiers.

For regeneration, Rehearsim separates locked or researcher-placed quotes from the candidates supplied to C13--C15. Locked theme titles accompany the remaining labels as already-used columns. If C14 returns one of those titles, application code merges its quote identifiers into the preserved theme by theme identifier. Split and merge operations carry the lock state forward, and unlocking a quote returns it to the candidate pool.

The API constructs C17 context from a fixed whitelist containing the study description, target population, project stage, and frozen interview questions. Transcript, quote-note, and theme-board fields are absent from the call. The interface displays the leading prompt-quality observation and reports the total number returned.

Each staged revision stores a target, before and after values, and a rationale. Applying the tray closes iteration $n$ and creates new question nodes for iteration $n+1$; carried questions receive \texttt{DERIVED\_FROM} links to their predecessors. Edits, additions, deletions, and reordering are applied in one transaction. The roster is omitted from the fork, forcing participant construction to run against the revised inputs before another interview begins.

\subsection{Runtime, Versioning, and Reproducibility}
\label{app:system-runtime}

Long operations are represented as database-backed jobs with an exclusive attempt identifier, heartbeat, completion state, and error state. Every write reasserts the active attempt identifier, preventing a stale worker from committing after another attempt has taken ownership. Because job state and intermediate artifacts reside in Neo4j, the client can reconnect to an operation already in progress and rehydrate its completed result.

Structured model outputs use provider-side JSON schemas. Transient provider failures receive randomized exponential backoff, truncated structured outputs receive one enlarged-budget retry, and permanent request errors terminate the affected job. Participant-specific calls preserve their assigned provider family across retries; exhaustion ends the job with that assignment recorded.

The prepare snapshot freezes fifteen prompt versions and the participant--claim relation version alongside the iteration inputs. Cache identity includes the source fields that cross each prompt boundary, so changing a subgroup screener, evidence selection, stimulus, or prompt version invalidates the affected derived artifact. A deterministic mock mode replaces retrieval and model calls with fixtures for end-to-end offline execution. Together, the frozen snapshot, artifact-level provenance, stable merges, and mock mode support replaying the workflow and auditing how an iteration was produced.

\section{User Study Materials}
\label{app:study-materials}

\subsection{Participants}

\label{sec:participants}
\begin{table*}
\centering
\small
\caption{Participant characteristics (N=25). Protocol-design experience is self-rated on a 1--7 scale.}
\label{tab:participants}
\begin{tabular}{llllclcl}
\toprule
ID & Role & Research area & Studies led & Exp. & GenAI freq. & Gender & Age \\
\midrule
R1 & Postdoc & HCI & 6+ & 7 & Daily & Female & 34 \\
R2 & Master's student & HCI & 3-5 & 2 & Daily & Male & 25 \\
R3 & PhD student & HCI & 6+ & 5 & Daily & Male & 29 \\
R4 & PhD student & HCI, User-Centered Design & 1-2 & 4 & Daily & Female & 27 \\
\midrule
P1 & PhD student & Usable Security \& Privacy, Social Science & 1-2 & 3 & Daily & Male & 27 \\
P2 & PhD student & HCI & 1-2 & 3 & Daily & Female & 26 \\
P3 & PhD student & HCI & 6+ & 7 & Daily & Male & 31 \\
P4 & PhD student & Health AI & 3-5 & 5 & Several/week & Female & 28 \\
P5 & PhD student & HCI, Education & 1-2 & 5 & Daily & Male & 27 \\
P6 & PhD student & HCI & 6+ & 6 & Daily & Male & 30 \\
P7 & PhD student & HCI & 1-2 & 4 & Daily & Male & 26 \\
P8 & PhD student & HCI & 3-5 & 7 & ~Once/week & Male & 29 \\
P9 & Master's student & Info Science, HCI & 1-2 & 6 & Daily & Female & 24 \\
P10 & PhD student & Health Communication & 3-5 & 5 & Daily & Female & 28 \\
P11 & Faculty & Business and HCI & 6+ & 7 & Daily & Male & 42 \\
P12 & Industry/UX researcher & HCI & 3-5 & 6 & Daily & Male & 33 \\
P13 & PhD student & HCI, Cultural Informatics, XR & 3-5 & 4 & Daily & Female & 27 \\
P14 & PhD student & AI EdTech Design, Co-design & 3-5 & 5 & Several/week & Female & 29 \\
P15 & PhD student & HCI \& Design & 3-5 & 6 & Daily & Female & 30 \\
P16 & Faculty & Design & 6+ & 4 & Several/week & Female & 45 \\
P17 & Faculty & Social Science; Recreation, Sport \& Tourism & 6+ & 7 & Daily & Female & 47 \\
P18 & PhD student & Game Studies/Historic Preservation & 1-2 & 3 & Daily & Female & 26 \\
P19 & Postdoc & HCI & 6+ & 6 & Daily & Male & 35 \\
P20 & PhD student & Education & 3-5 & 3 & Daily & Female & 28 \\
P21 & Faculty & HCI, Computational Social Science, Communication & 3-5 & 6 & Daily & Female & 39 \\
\bottomrule
\end{tabular}
\end{table*}

\subsection{Survey}
\label{survey}
\subsubsection{Background Survey}

\newlist{surveyQ}{enumerate}{1}
\setlist[surveyQ]{label={(\arabic*)},leftmargin=*}
\newlist{surveyOpt}{enumerate}{1}
\setlist[surveyOpt]{label={(\alph*)},leftmargin=*}

\begin{enumerate}
\footnotesize
\item[] \textbf{\small{Section 1. Research Background}}
\item \textbf{Which best describes your current role? }
    \begin{enumerate}
        \item Master's student
        \item PhD student
        \item Postdoc
        \item Faculty
        \item Research staff
        \item Industry/UX researcher
        \item Other
    \end{enumerate}
\item \textbf{Primary field/discipline (e.g., HCI, social science, education, health, design)} (Text field)
\item \textbf{What type of HCI researcher do you self identify as (e.g. qualitative researcher, mixed-method researcher...)} (Text field)
\item \textbf{How many qualitative studies have you led or co-led? }
    \begin{enumerate}
        \item 0
        \item 1-2
        \item 3-5
        \item 6+
    \end{enumerate}
\item \textbf{If you have published qualitative research papers, which conferences were they published at, and how many have you published in total?} (Text field)
\item \textbf{What qualitative methods you have applied in your prior work, select all that apply?}
    \begin{enumerate}
        \item Interviews
        \item Focus groups
        \item Observation or ethnographic methods
        \item Think-aloud or diary studies
        \item Workshops or co-design sessions
        \item Open-ended surveys
        \item Qualitative data analysis, such as thematic analysis or coding
        \item I have not previously used qualitative methods
        \item Others (text-field)
    \end{enumerate}
\item \textbf{What best describes your training in qualitative research methods? (Select all that apply)?}
    \begin{enumerate}
        \item Formal coursework
        \item Workshops/mentorship
        \item Self-taught
        \item None
        \item Others (text-field)
    \end{enumerate}

\end{enumerate}

\begin{enumerate}
\footnotesize
\item[] \textbf{\small{Section 2. Qualitative Research Skills}}\\
Please rate the following statements on a scale from 1 (Strongly
Disagree) to 7 (Strongly Agree)
\item Experience designing interview or qualitative study protocols?

\item Experience conducting interviews, focus groups, or qualitative fieldwork?

\item Experience with thematic analysis or qualitative coding?

\item Confidence designing a qualitative protocol on your own?

\item How much mental effort do you typically invest in designing a qualitative research protocol? \end{enumerate}

\begin{enumerate}
\footnotesize
\item[] \textbf{\small{Section 3. AI Experience and Attitudes}}

\item \textbf{How often do you use Generative AI tools (e.g. ChatGPT, Claude, Gemini, Copilot) in your research workflow?} 
    \begin{enumerate}
        \item Daily
        \item Several times a week
        \item About once a week
        \item A few times a month
        \item Rarely
        \item Never
    \end{enumerate}
    
\item \textbf{How would you describe your overall experience and proficiency with generative AI tools?}

\item \textbf{Have you ever used any AI tools to assist with any stage of your qualitative research?} 
    \begin{enumerate}
        \item Yes
        \item No
    \end{enumerate}

\item \textbf{If yes, which of the following qualitative research tasks have you used AI to assist with? (Select all that apply)?} 
    \begin{enumerate}
        \item Brainstorming research questions
        \item Designing or refining study protocols/interview guides
        \item Generating or refining interview questions
        \item Simulating participants, users, or personas
        \item Analyzing qualitative data, such as coding or theme identification
        \item Writing or editing manuscript
        \item literature review
        \item Others (text-field)
    \end{enumerate}

\item \textbf{I am comfortable using AI-generated materials to inform revisions to a qualitative research protocol.} (scale from 1 (Strongly
Disagree) to 7 (Strongly Agree))

\item \textbf{How concerned are you about using AI-generated simulated participant responses to inform qualitative research design?} (scale from 1 (Strongly
Disagree) to 7 (Strongly Agree))

\item \textbf{What are the concerns or risks? } (Text field)

\end{enumerate}

\subsubsection{Survey 1}

\begin{enumerate}
\footnotesize
\item[] \textbf{\small{Research Context Familiarity}}\\
Please rate the following statements on a scale from 1 (Strongly
Disagree) to 7 (Strongly Agree)
\item I am familiar with the research context of this protocol.
\item I am familiar with the target population of this protocol.
\end{enumerate}
\begin{enumerate}
\footnotesize
\item[] \textbf{\small{Quality of the Initial Interview Protocol}}\\
Please rate the following statements on a scale from 1 (Strongly Disagree) to 7 (Strongly Agree).

\item Scope: I am clear about which settings and situations are within this study~\cite{dejonckheere2019semistructured}.

\item Scope: I am clear about who should be included in my target population~\cite{dejonckheere2019semistructured}.

\item Coverage: My current protocol covers the key aspects of the research question without leaving major gaps~\cite{castillomontoya2016preparing,kallio2016systematic}.

\item Clarity: My current protocol includes questions that are clear, specific, and easy for participants to understand~\cite{castillomontoya2016preparing,yeong2018interview}.

\item Depth: My current protocol would likely elicit rich, detailed, and reflective responses from participants~\cite{castillomontoya2016preparing,kallio2016systematic}.

\item Breadth: My current protocol captures a range of relevant perspectives or experiences from participants~\cite{kallio2016systematic}.

\item Rigorous: My current protocol is methodologically rigorous~\cite{kallio2016systematic}.

\item Feasibility: My current protocol is feasible to use in a real pilot study~\cite{castillomontoya2016preparing,kallio2016systematic}.
\end{enumerate}
\begin{enumerate}
\footnotesize
\item[] \textbf{\small{Protocol Assumption Awareness}}\\
Please rate the following statements on a scale from 1 (Strongly
Disagree) to 7 (Strongly Agree)

\item I can identify assumptions my current protocol makes about participants’ experiences, knowledge, or circumstances.
\item I can identify gaps or missing contexts in my current protocol.
\item I can identify questions in my current protocol that may be vague, leading, or incomplete.
\end{enumerate}
\begin{enumerate}
\footnotesize
\item[] \textbf{\small{Revision Readiness}}\\
Please rate the following statements on a scale from 1 (Strongly
Disagree) to 7 (Strongly Agree)
\item I know concrete ways to revise my current protocol.
\item I can prioritize which parts of my current protocol need revision.
\item I feel confident revising my protocol.

\end{enumerate}

\subsubsection{Survey 2 \& 3}
\begin{enumerate}
\footnotesize
\item[] \textbf{\small{Quality of the Revised Protocol}}\\
Please rate the following statements on a scale from 1 (Strongly
Disagree) to 7 (Strongly Agree)
\item Scope: I am clear about which settings and situations are within the scope of this study.
\item Scope: I am clear about who should be included in my target population.
\item Coverage: My revised protocol covers the key aspects of the research question without leaving major gaps.
\item Clarity: My revised protocol includes questions that are clear, specific, and easy for participants to understand.
\item Depth: My revised protocol would likely elicit rich, detailed, and reflective responses from participants.
\item Breadth: My revised protocol captures a range of relevant perspectives or experiences from the target population.
\item Rigorous: My revised protocol is methodologically rigorous.
\item Feasibility: My revised protocol is feasible to use in a real pilot study.
\end{enumerate}

\begin{enumerate}
\footnotesize
\item[] \textbf{\small{Protocol Assumption Probing}}\\
Please rate the following statements on a scale from 1 (Strongly
Disagree) to 7 (Strongly Agree)
\item After using the tool, I can identify assumptions my initial protocol makes about participants’ experiences, knowledge, or circumstances.
\item After using the tool, I can identify gaps or missing contexts in my initial protocol.
\item After using the tool, I can identify questions in my initial protocol that may be vague, leading, or incomplete.
\end{enumerate}

\begin{enumerate}
\footnotesize
\item[] \textbf{\small{Revision Process Support}}\\
Please rate the following statements on a scale from 1 (Strongly
Disagree) to 7 (Strongly Agree)
\item After using the tool, I know concrete ways to revise my protocol.

\item After using the tool, I can prioritize which parts of my protocol need revision.
\item After using the tool, I feel confident revising my protocol.
\item I would use this tool before piloting a qualitative study with real participants.
\end{enumerate}

\begin{enumerate}
\footnotesize
\item[] \textbf{\small{Cognitive Load}}\\
Please rate the following statements on a scale from 1 (Strongly
Disagree) to 7 (Strongly Agree)
\item How much mental effort did you invest in revising the interview protocol using the system?

\item How mentally demanding was it to understand and operate the system?
\item How irritated, stressed, or annoyed did you feel using the system?
\end{enumerate}
\begin{enumerate}
\footnotesize
\item[] \textbf{\small{Critical Use of Simulated Responses}}\\
Please rate the following statements on a scale from 1 (Strongly
Disagree) to 7 (Strongly Agree)
\item The simulated responses were consistent with the participant profiles they were based on.
\item The simulated responses resembled what people in the target population might plausibly say.
\item The simulated responses were useful as provisional materials for critique.
\item The simulated responses surfaced reactions I had not anticipated.
\item I could tell which simulated responses were useful for protocol revision and which should be ignored.
\item I treated the simulated responses as evidence about what real participants would say.
\item I would replace real human participants with the simulated participants.
\item I am comfortable using AI-generated materials to inform revisions to a qualitative research protocol.
\item How concerned are you about using AI-generated simulated participant responses to inform qualitative research design?
\end{enumerate}

\begin{enumerate}
\footnotesize
\item[] \textbf{\small{System Features}}

\item \textbf{Feature 1: Co-design literature-grounded simulated participant profiles.}\\
Please rate the following statements on a scale from 1 (Strongly
Disagree) to 7 (Strongly Agree) 
    \begin{enumerate}
        \item I enjoyed using this feature.
        \item I found this feature useful for improving my protocol design.
        \item I trusted the suggestions and decisions provided by it.
        \item This feature helped me think more critically about my research protocol.
        \item This feature contributed to revising my protocol. 
    \end{enumerate}

\item \textbf{Feature 2: Reviewing simulated results and practicing follow-up questions.}\\
Please rate the following statements on a scale from 1 (Strongly
Disagree) to 7 (Strongly Agree) 
    \begin{enumerate}
        \item I enjoyed using this feature. 	
        \item I found this feature useful for improving my protocol design. 	
        \item I trusted the suggestions and decisions provided by it.  	
        \item This feature helped me think more critically about my research protocol. 	
        \item This feature contributed to revising my protocol.
    \end{enumerate}

\item \textbf{Feature 3: Thematic analysis dashboard.} \\
Please rate the following statements on a scale from 1 (Strongly
Disagree) to 7 (Strongly Agree)
    \begin{enumerate}
        \item I enjoyed using this feature.		
        \item I found this feature useful for improving my protocol design. 	
        \item I trusted the suggestions and decisions provided by it.	
        \item This feature helped me think more critically about my research protocol. 	
        \item This feature contributed to revising my protocol. 
    \end{enumerate}
    
\end{enumerate}







\end{document}